\documentclass[pra,twocolumn,superscriptaddress,floatfix]{revtex4}
\usepackage{graphicx}
\usepackage[dvips]{epsfig}
\usepackage{amssymb,amsmath,latexsym,bm}
\usepackage{txfonts} 
\usepackage{verbatim}

\renewcommand{\vec}[1]{{\bm{\mathrm{#1}}}}
\newcommand{\vhat}[1]{\hat{\bm{\mathrm{#1}}}}
\newcommand{\ocite}[1]{Ref.~[\onlinecite{#1}]}
\let\epsilon\varepsilon

\begin{document}
\title{Theory of Kondo suppression of spin polarization in nonlocal spin valves}
\author{K.-W. Kim}%
\thanks{K.-W. Kim and L. O'Brien contributed equally to this work.}
\affiliation{Center for Nanoscale Science and Technology, National Institute
of Standards and Technology, Gaithersburg, Maryland 20899, USA}%
\affiliation{Maryland NanoCenter, University of Maryland, College Park,
Maryland 20742, USA}%
\affiliation{Institut f\"{u}r Physik, Johannes Gutenberg Universit\"{a}t Mainz, Mainz 55128, Germany}%
\author{L. O'Brien}%
\thanks{K.-W. Kim and L. O'Brien contributed equally to this work.}
\affiliation{Department of Chemical Engineering and Materials Science,
University of Minnesota, Minnesota, 55455 USA}%
\affiliation{Thin Film Magnetism, Cavendish Laboratory, University of Cambridge, CB3 0HE, UK}%
\author{P. A. Crowell}%
\affiliation{School of Physics and Astronomy,
University of Minnesota, Minnesota 55455, USA}%
\author{C. Leighton}%
\email{leighton@umn.edu}%
\affiliation{Department of Chemical Engineering
and Materials Science,
University of Minnesota, Minnesota 55455, USA}%
\author{M. D. Stiles}%
\email{mark.stiles@nist.gov}%
\affiliation{Center for Nanoscale Science and Technology, National Institute
of Standards and Technology, Gaithersburg, Maryland 20899, USA}%
\date{\today}

\begin{abstract}
We theoretically analyze contributions from the Kondo effect to the spin
polarization and spin diffusion length in all-metal nonlocal spin valves. Interdiffusion of ferromagnetic atoms into the normal metal
layer creates a region in which Kondo physics plays a significant role, giving discrepancies between experiment and existing theory. We
start from a simple model and construct a modified spin drift-diffusion
equation which clearly demonstrates how the Kondo physics not only
suppresses the electrical conductivity but even more strongly reduces the
spin diffusion length. We also present an explicit expression for the
suppression of spin polarization due to Kondo physics in an
illustrative regime. We compare this theory to previous experimental data
to extract an estimate of the Elliot-Yafet probability for Kondo spin flip scattering of 0.7 $\pm$ 0.4, in good agreement with the value of 2/3 derived in the original theory of Kondo.
\end{abstract}


\maketitle

\section{Introduction\label{Sec:Intro}}

Pure spin currents, devoid of charge current flow, are now routinely generated in metals-based systems~\cite{Bass07JPCM} via a number of techniques, including the use of thermal gradients~\cite{Uchida08Nat}, the spin Hall effect~\cite{Valenzuela06Nat}, spin pumping~\cite{Tserkovnyak05RMP}, and nonlocal spin injection~\cite{Johnson85PRL}, each method providing a unique insight into spin relaxation. In particular, the ability to separate charge and spin currents using the nonlocal spin valve~\cite{Johnson85PRL,Jedema01Nat}, thereby circumventing difficulties interpreting `local' spin valve measurements, makes it one of the most unambiguous techniques for probing spin transport. This geometry is especially useful at the nanoscale, where isolating the factors affecting spin accumulation, diffusion, and relaxation, both within the bulk and across interfaces, represents a pressing problem~\cite{Garzon05PRL,Godfrey06PRL,Kimura07PRL,Yang08NP,Bakker10PRL,Mihajlovic10PRL,Slachter10NP,Fukuma11NM,Niimi13PRL}. Indeed, examining the role of specific defects in relaxing spins in metals at this length scale -- including interfaces, grain boundaries, and magnetic and highly spin-orbit coupled impurities -- will be critical for realizing future low resistance-area-product spintronic devices, e.g. current perpendicular-to-plane giant magnetoresistance sensors~\cite{Gijs97AIP}.

\begin{figure}
\includegraphics[width=8.6cm]{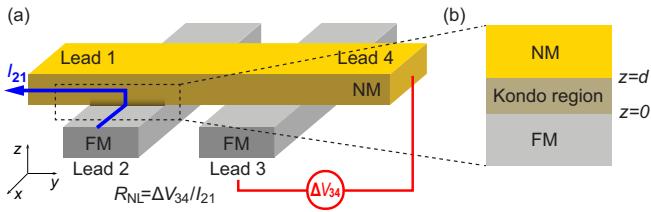}
\caption{%
(color online) (a) Geometry of a nonlocal spin valve, consisting of a normal metal (NM) channel and two ferromagnetic (FM) contacts. When an
electrical current $I_{21}$ is applied (blue line), a nonequilibrium spin
accumulation develops, giving rise to a finite voltage difference $\Delta V_{34}$ and nonlocal resistance $R_{\rm NL}=\Delta V_{34}/I_{21}$. (b) Schematic of
the model used to determine the suppression of spin polarization at a ferromagnet/nonmagnet interface ($z=0$) due to interdiffused ferromagnetic atoms over a characteristic
length scale $d$, which we term the Kondo region.
}%
\label{Fig1:NLSV}
\end{figure}

A nonlocal spin valve consisting of a normal metal channel connected by two ferromagnetic contacts is illustrated in Fig.~\ref{Fig1:NLSV}(a): the injected current $I_{21}$
generates a spin accumulation at the interface between the nonmagnet and ferromagnet (Lead 2). This accumulation diffuses in both directions down the channel causing a pure spin current to flow towards Lead 1, which decays on a characteristic spin diffusion
length, $l_{N}^{\rm sf}$. The remaining spin population reaching Lead 3 generates a nonlocal voltage difference $
V_{34}$ between the ferromagnetic contact and channel and therefore a nonlocal resistance, $ R_{\rm NL}=\Delta V_{34}/I_{21}$. The sign of this resistance depends on the relative orientation of the two ferromagnets, and so by applying a magnetic field to alternate the ferromagnetic contact magnetization from parallel to antiparallel, a nonlocal spin signal, $\Delta R_{\rm NL}$, is measured, directly related to the magnitude of the spin accumulation under the contact.

In relatively simple all-metal nonlocal spin valves (e.g., Ni$_{80}$Fe$_{20}$/Cu) that are fabricated from nominally high-purity materials, the standard theory of spin drift-diffusion developed by Valet and Fert~\cite{Valet93PRB}, combined with the Elliott-Yafet spin relaxation mechanism~\cite{Elliott54PR,Andreev59JETP,Yafet63Book}, which predominates in light metals, dictates that $l_N^{\rm sf}$, the spin accumulation, and therefore $\Delta R_{\rm NL}$ should monotonically increase as temperature $T$ decreases. Surprisingly, however, $\Delta R_{\rm NL}$ is widely found to anomalously \textit{decrease} at low $T$ in Ni$_{80}$Fe$_{20}$/Cu, Fe/Cu and Co/Cu nonlocal spin valves~\cite{Kimura08PRL,Casanova09PRB,Otani11PTAMPES,Erekhinsky12APL,Kimura12PRB,Zou12APL,Villamor13PRB,Brien14NC,Batley15PRB,Brien16PRB}, even when the resistivity of the normal metal and the ferromagnet, $\rho_N(T)$ and $\rho_F(T)$, are found to continuously decrease on cooling.

	Consensus is emerging that this unexpected reduction of $\Delta R_{\rm NL}$ at low $T$ is due to spin relaxation at dilute magnetic impurities~\cite{Zou12APL,Villamor13PRB,Batley15PRB}, with recent results demonstrating that a manifestation of the Kondo effect is at the heart of the suppression~\cite{Brien14NC,Brien16PRB}. The Kondo effect~\cite{Kondo64PTP} arises in metals with dilute magnetic impurities, as a result of $s$-$d$ exchange between the conduction electrons and virtual bound impurity states. This exchange results in an additional higher order contribution to the scattering cross section, proportional to $\log T$, which can dominate in otherwise highly pure metals at low $T$. In charge transport, the classic signature of the Kondo effect is an increase in the conduction electron scattering rate at low $T$, resulting in a minimum in resistance (maximum in conductance) and a logarithmic increase in $\rho(T)$ about a characteristic temperature $T_{K}$~\cite{Domenicali61JAP}. Similarly, for spin transport the additional (spin-flip) scattering was recently found to efficiently relax the spin accumulation, suppressing $\Delta R_{\rm NL}$ with what is also observed to be a log $T$ dependence~\cite{Brien14NC}. This occurs even for nonmagnetic channels that are largely impurity-free throughout the bulk, due to inevitable interdiffusion at the ferromagnet/nonmagnet interface. The situation is schematically depicted in Fig.~\ref{Fig1:NLSV}(b), where interdiffusion creates a region with `high' levels of ferromagnetic impurities (on average $\approx 100$'s $\mu$mol/mol~\footnote{$\mu$mol/mol is equivalent to `parts per million'}) which rapidly relax the injected spins at the interface, reducing the effective polarization of the bias current~\cite{Brien14NC}. To this point, a direct quantitative link has already been established between the degree of interdiffusion and magnitude of Kondo suppression in nonlocal spin valves~\cite{Brien16PRB}. Reciprocally, the disruption of the Kondo singlet through the injection of sufficiently large spin currents has also been investigated~\cite{Taniyama16, Taniyama03PRL}. Since there are no magnetic impurities far away from the interface, spin diffusion in the bulk is described by the spin drift-diffusion equation in the Valet-Fert theory, and a measure of $\rho_N(T)$ yields no indication of the Kondo effect. Naturally, in devices where impurity levels are sufficiently high throughout the nonmagnet, either due to intentional doping~\cite{Batley15PRB}, source contamination~\cite{Villamor13PRB}, or contamination during deposition~\cite{Zou12APL}, the effects of Kondo scattering can equally be found to enhance spin relaxation in the bulk of the channel, thereby reducing $l_{N}^{\rm sf}$. Despite this growing body of experimental work, a complete theoretical treatment of the effect remains outstanding; a description of the suppression of the spin polarization near the interface due to the Kondo effect is therefore the aim of this work.

In this paper, we start from the Boltzmann equation and follow the Valet-Fert theory~\cite{Valet93PRB} to construct a modified spin drift-diffusion equation
which is valid in the presence of dilute magnetic impurities. While the Valet-Fert theory is developed for $T=0$, we allow for nonzero temperature to extract
Kondo contributions. Then, we project our theory to a low-temperature regime, keeping the additional Kondo contributions. Using the modified spin
drift-diffusion equation, we compare our theory to experimental data to extract an estimate of the Elliot-Yafet parameter for Kondo spin relaxation. This is found to be in very good agreement with the value originally proposed by Kondo~\cite{Kondo64PTP}.

The paper is organized as follows. In Sec.~\ref{Sec:Result}, we develop a
theory describing suppression of the spin polarization at the interface. We
first present the theory without derivation in Sec.~\ref{Sec:Result-A} and
then we demonstrate that the spin polarization at the interface has a maximum at a finite temperature. In Sec.~\ref{Sec:Fit}, we compare the theory to our
experimental data~\cite{Brien16PRB}. In Sec.~\ref{Sec:Theory}, we present mathematical details which are referred to in Sec.~\ref{Sec:Result}. Finally, in
Sec.~\ref{Sec:Summary} we summarize the paper.

\section{interfacial Kondo effect\label{Sec:Result}}

\subsection{Modification of the electrical conductivity and the spin diffusion length due to the Kondo effect\label{Sec:Result-A}}

Starting from antiferromagnetic exchange coupling between conduction
electrons and dilute magnetic impurities, Kondo~\cite{Kondo64PTP} showed that
electrical conductivity in metals is suppressed at low temperature. This is equivalent to suppression of the momentum relaxation time:
\begin{equation}
\frac{1}{\tilde{\tau}_N}=\frac{1}{\tau_N}+\frac{1}{\tau_K^{\rm eff}},\label{Eq:tauNF relation}
\end{equation}
where $\tau_N$ is the momentum relaxation time without dilute magnetic
impurities in the normal metal at the Fermi level and $\tilde{\tau}_N$ is the modified momentum relaxation time in the presence of the Kondo effect.
$\tau_K^{\rm eff}$ is the effective Kondo relaxation time, for which the
explicit expression is given below in Eq.~(\ref{Eq:Kondo rate}). $\tau_K^{\rm eff}$ has a logarithmic temperature dependence, and thus it can be comparable to or
even dominate $\tau_N$ at low temperature for very low impurity concentrations.

In addition to suppressing the scattering time (increasing the scattering rate), as in Eq.~(\ref{Eq:tauNF
relation}), the dilute magnetic impurities also suppress the spin
relaxation time $\tau_N^{\rm sf}$.
\begin{equation}
\frac{1}{\tilde{\tau}_N^{\rm sf}}=\frac{1}{\tau_N^{\rm sf}}+\frac{\eta}{\tau_K^{\rm eff}},\label{Eq:tauNsf relation}
\end{equation}
where $\tilde{\tau}_N^{\rm sf}$ is the modified spin relaxation time due to
the Kondo effect. Here, $\eta$ is the spin-flip probability during each Kondo scattering event. The proportionality between the change in the momentum relaxation rate ($1/\tau_K^{\rm eff}$) and the spin relaxation rate ($1/\tau_K^{\rm eff}$) is similar to that found for the Elliot-Yafet scattering mechanism. For Elliot-Yafet scattering, the contribution to the spin-flip scattering rate is given by $1/\beta\tau$ where $1/\tau$ is the contribution to the momentum scattering rate and $\beta$ is the Elliot-Yafet parameter. Thus, $\eta$ is the inverse of the Elliot-Yafet parameter for spin relaxation from Kondo impurities. The value of $\eta$ is determined by the geometry of the
Fermi surface. For the spherical Fermi surfaces that we consider here, $\eta=2/3$, as shown in Sec.~\ref{Sec:Theory}. We note that in \ocite{Batley15PRB}, the spin-flip
probability is claimed to be around 0.3 based on a semiclassical argument~\cite{Fert95PRB}. Strictly, however, the semiclassical argument does not apply for the higher
order interactions giving rise to the Kondo physics.

One remark on our notation is in order. Generally speaking $\tau_N$ and
$\tau_N^{\rm sf}$ are $k$-dependent functions. But here, we drop the $k$
dependence and take the values only at the Fermi level for simplicity.
In Sec.~\ref{Sec:Theory}, we restore the $k$ dependence for the
derivation. In particular, $\tau_N(k_N^F)$ and $\tau_N^{\rm sf}(k_N^F)$ in
Sec.~\ref{Sec:Theory} are respectively $\tau_N$ and $\tau_N^{\rm sf}$ here,
where $k_N^F$ is the Fermi wavevector in the normal metal.

The changes in the relaxation times above imply changes in the electrical
conductivity and the spin diffusion length. In Sec.~\ref{Sec:Theory}, we show
that the Valet-Fert theory for the spin drift-diffusion equation still holds
even in the presence of dilute magnetic impurities, once we impose
Eqs.~(\ref{Eq:tauNF relation}) and (\ref{Eq:tauNsf relation}). The Valet-Fert
theory provides links between quantities in the Boltzmann equation (such as
relaxation times) and quantities in the drift-diffusion equation (such as the
electrical conductivity and spin diffusion length) as given below in
Eqs.~(\ref{Eq:sigma to tau}) and (\ref{Eq:l to tau}). Up to first order
in the Kondo rate, the modified conductivity and spin diffusion length
are
\begin{align}
\tilde{\sigma}_N&= \sigma_N\left(1-\frac{\tau_N}{\tau_K^{\rm eff}}\right),\label{Eq:renormalization sigma}\\
{\tilde{l}_N}^2&= {l_N}^2\left(1-\frac{\tau_N+\eta\tau_N^{\rm sf}}{\tau_K^{\rm eff}}\right).\label{Eq:renormalization l}
\end{align}
Equation~(\ref{Eq:renormalization sigma}) corresponds to Kondo's original
work, i.e. the conductivity reduction due to the Kondo effect.
Equation~(\ref{Eq:renormalization l}) is the spin counterpart of the original
Kondo effect, a central aspect of this paper.

At this point it is worth noting that the Kondo effect can affect the spin diffusion length much more dramatically
than it does the conductivity, since $(\tau_N^{\rm sf})^{-1}$ is usually much
smaller than $\tau_N^{-1}$. For example, $\tau_N^{\rm sf}/\tau_N\approx10^3$
in \ocite{Bass07JPCM}. Thus it is possible that $1-{\tilde{l}_N}^2/{l_N}^2$ is noticeable even though $1-\tilde{\sigma}_N/\sigma_N$ is negligible.

\subsection{Suppression of the spin polarization at the interface\label{Sec:Result-B}}

We now solve the spin drift-diffusion equation with the modified quantities in
Eqs.~(\ref{Eq:renormalization sigma}) and (\ref{Eq:renormalization l}) at the interface. Our model is illustrated in Fig.~\ref{Fig1:NLSV}(b). We consider a
ferromagnet $(z<0)$/nonmagnet $(z>0)$ interface at $z=0$. Near the interface, interdiffused
ferromagnetic atoms create a region in which dilute magnetic impurities are
present. For illustration, we assume that the impurity concentration
is constant over $0<z<d$ and suddenly drops to zero at $z=d$. We term the region
$0<z<d$ the Kondo region.

The spin drift-diffusion equation is given by the set of equations below~\cite{Valet93PRB}.
\begin{align}
\frac{e}{\sigma_s(z)}\partial_zj_s(z)&=\frac{\mu_s(z)-\mu_{-s}(z)}{{l_s}^2(z)},\label{Eq:V-F eq1}\\
\partial_z\mu_s(z)&=\frac{e}{\sigma_s(z)}j_s(z),\label{Eq:V-F eq2}
\end{align}
where $s=\pm$ denotes the spin majority and minority bands, $e$ is the
electron charge, $j_s(z)$, $\mu_s(z)$, $\sigma_s(z)$, and $l_s(z)$ are
respectively the current expectation value, the chemical potential, the
electrical conductivity, and the spin diffusion length of the spin $s$ band
at position $z$. $\sigma_s$ and $l_s$ are parameters treated as
position independent in most cases. We explicitly retain the position dependence to emphasize the dependence of the parameters on the
regions; $z<0$, $0<z<d$, and $z>d$. More explicitly, the parameters in each
region are given by
\begin{equation}
(\sigma_s(z),l_s(z))=\left\{\begin{array}{cl} (\sigma_{s,F},l_{s,F})&\mathrm{for~}z<0,\\
(\tilde{\sigma}_N,\tilde{l}_N)&\mathrm{for~}0<z<d,\\
(\sigma_N,l_N)&\mathrm{for~}z>d.\end{array}\right.\label{Eq:sigma and l with Kondo}
\end{equation}
Here the subscripts $N$ and $F$ refer to the normal metal and the
ferromagnetic metal and the tildes refer to the Kondo region~%
\footnote{We emphasize that each region is specified by subscripts. Note that
the \emph{superscript} $F$ that appears in $k_s^F$ for instance refers to ‘at
the Fermi surface’.}. %
Since the physical parameters in the normal metal do not have
spin dependence, we drop the subscript $s$ for $z>0$.

\begin{figure}
\includegraphics[width=8.6cm]{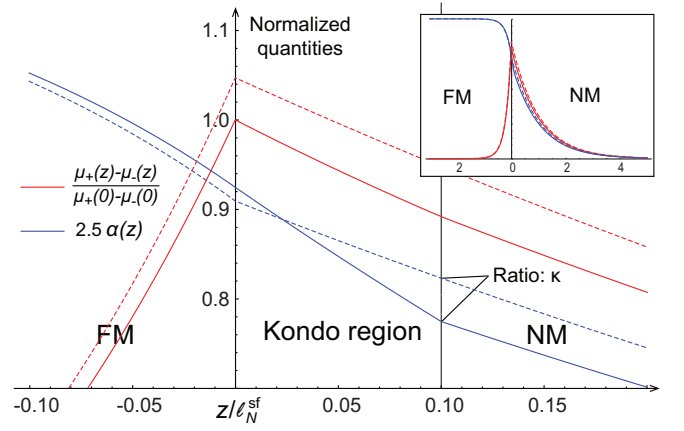}
\caption{(color online) Profiles of the spin accumulation $\mu_+(z)-\mu_-(z)$ divided by its value at $z=0$ (red), and the current polarization $\alpha(z)=[j_+(z)-j_-(z)]/j_{\rm app}$ multiplied by a factor of 2.5 (blue) to allow it to be plotted on the same scale as the spin accumulation. The dashed lines denote the solutions without dilute magnetic impurities with the same normalization factors. $\kappa$ is defined as the ratio of $\alpha(d)$ for the case of finite magnetic impurity concentration to that when no magnetic impurities are present [Eq.~(\ref{Eq:kappa0})]. $\kappa$ defined by the spin accumulation and that defined by the spin current have the same value. The inset shows the profiles over a wider range
$z/l_N^{\rm sf}=-3$ to $5$. The parameters used here are
$\sigma_{\pm,F}=(0.1\pm0.05)\sigma_N$, $l_N^{\rm sf}=5l_F^{\rm sf}=10d$,
$\tilde{l}_N^{\rm sf}/l_N^{\rm sf}=0.9$, and $\tilde{\sigma}_N/\sigma_N=0.7$.
} \label{Fig3:Plot}
\end{figure}

The general solutions of Eqs.~(\ref{Eq:V-F eq1}) and (\ref{Eq:V-F eq2}) are
obtained in \ocite{Valet93PRB}. The spin accumulation $\mu_+-\mu_-$ decays
exponentially over the effective spin diffusion length defined by $(l_F^{\rm
sf})^{-2}={l_{+,F}}^{-2}+{l_{-,F}}^{-2}$ for $z<0$, $(\tilde{l}_N^{\rm
sf})^{-2}=2{\tilde{l}_N}^{-2}$ for $0<z<d$, and $(l_N^{\rm sf})^{-2}=2{l_N}^{-2}$ for $z>d$. To match the experimental situation of Refs.~\cite{Brien14NC} and~\cite{Brien16PRB}, we assume transparent interfaces at $z=0$ and $z=d$ where the spin
chemical potential and the currents are continuous. The solutions of
Eqs.~(\ref{Eq:V-F eq1}) and (\ref{Eq:V-F eq2}) in the given situation are
obtained in Appendix~\ref{Sec-A:BC}. Here, we simply present
spatial profiles of the spin accumulation $\mu_+(z)-\mu_-(z)$ and the spin
current $j_+(z)-j_-(z)$ for a set of parameters in Fig.~\ref{Fig3:Plot}. For clear presentation, we scale the quantities. In this case, $\mu_+ (z) -\mu_-(z)$ is normalized to its $z=0$ value, and $j_+ (z) - j_- (z)$ to $j_{\rm app}/2.5$. Here $j_{\rm app}=j_+(z)+j_-(z)$ is the applied charge current which is independent of position due to the conservation of electrical charge. The factor of 2.5 is simply to allow both quantities to be plotted on the same scale.

Figure~\ref{Fig3:Plot} clearly shows that there is suppression of the spin
polarization due to the Kondo region. Since there are no magnetic impurities for $z>d$, relaxation rates for $z>d$ are the same with and without the Kondo region. Thus, the spin
accumulation calculated at $z=d$ indicates the suppression of $\Delta R_{\rm
NL}$ due to the Kondo effect. To quantify this suppression, we analytically evaluate the ratio of the accumulation at $d$ in the presence and absence of a finite impurity concentration in the Kondo region (i.e. in the limit $\tau_K^{\rm eff}\to\infty$). Here we compute the following expression up to $\mathcal{O}((\tau_{K}^{\rm eff})^{-1})$:%
\begin{equation}
\kappa\equiv\frac{\mu_+(d)-\mu_-(d)}{\lim_{\tau_{K}^{\rm eff}\to\infty}[\mu_+(d)-\mu_-(d)]}\equiv\frac{\alpha(d)}{\lim_{\tau_{K}^{\rm eff}\to\infty}[\alpha(d)]},\label{Eq:kappa0}
\end{equation}
where $\alpha(z)$ is defined as the current polarization, $\alpha(z)=[j_+(z) - j_-(z)]/j_{\rm app}$~\footnote{One can verify explicitly that $\kappa$
has the same value for both definitions.}. At this point it is worth emphasising that, since it is a quantitative indication of the strength of Kondo suppression, the suppression ratio $\kappa$ represents a key parameter of this work. Furthermore, as $\alpha$ is directly measurable, $\kappa$ uniquely represents an experimentally accessible spin transport parameter with which to compare theory and measurement. In evaluating Eq.~(\ref{Eq:kappa0}) we retain only terms up to $\mathcal{O}(d)$, assuming that $d$ is much shorter
than the effective spin diffusion length. After straightforward but tedious algebra, we obtain
\begin{equation}
\kappa=1-d\left[\frac{1+2\eta(\tau_N^{\rm sf}/\tau_N)}{l_N^{\rm sf}}-\frac{2\eta(\tau_N^{\rm sf}/\tau_N)}{l_N^{\rm
sf}+\frac{\rho_F}{(1-{\alpha_{\rm FM}}^2)\rho_N}l_F^{\rm sf}}\right]\frac{\rho_K}{\rho_N},\label{Eq:Kappa3}
\end{equation}
where $\rho_K$ is the Kondo contribution to the resistivity,
$\rho_N=(2\sigma_N)^{-1}$ is the electrical resistivity of the normal metal, and $\rho_F=(\sigma_{+,F}+\sigma_{-,F})^{-1}$ is the electrical resistivity of the ferromagnet. From the Drude model, $\rho_K/\rho_N=\tau_N/\tau_K^{\rm eff}$. $\alpha_{\rm FM}=\alpha(z=-\infty)$, the current polarization far away from the interface, is a material parameter determined by the conductivity polarization $(\sigma_{+,F}-\sigma_{-,F})/(\sigma_{+,F}+\sigma_{-,F})$. The advantage of
writing Eq.~(\ref{Eq:Kappa3}) in terms of $\rho_K$ instead of
$\tau_K^{\rm eff}$ is that we can avoid the original Kondo expression
$\sim\log T$, which diverges at low temperature, and instead use the phenomenological
expression for $\rho_K$ suggested by Goldhaber-Gordon~\cite{Goldhaber-Gordon98PRL} [Eq.~(\ref{Eq:rhoGG})], which is known to work well for a wide range of
temperatures~\cite{Parks07PRL,vanderWiel00Sci}. If $\tau_N^{\rm sf}\gg\tau_N$, $1+2\eta(\tau_N^{\rm
sf}/\tau_N)\approx 2\eta(\tau_N^{\rm sf}/\tau_N)$, thus one can verify that
$1-\kappa$ is proportional to $(\tau_N^{\rm sf}/\tau_N)$, which is on the order of $10^3$ for Cu. Such a large factor shows that the spin diffusion length is indeed a good tool to observe the Kondo effect, as discussed in Sec.~\ref{Sec:Result-A}.

For an order-of-magnitude estimate of the suppression ratio $1-\kappa$ we take: $l_N^{\rm sf}\approx
500~\mathrm{nm}$ and $l_F^{\rm sf}\approx10~\mathrm{nm}$~\cite{Bass07JPCM}, with
$\rho_F/(1-{\alpha_{\rm FM}}^2)\rho_N\approx10$ and
$2\eta\tau_N^{\rm sf}/\tau_N\approx10^3$. With $\rho_K/\rho_N\approx0.01$, $1-\kappa$ is around a percent
for $d=1~\mathrm{nm}$ and is proportional to $d$. This crude
order-of-magnitude estimation is comparable to our previous experiments~\cite{Brien14NC,Brien16PRB}.

\section{Detailed comparison with experimental data\label{Sec:Fit}}

The suppression of the spin polarization $\alpha$ [quantified by $1-\kappa$ in Eq.~(\ref{Eq:Kappa3})] has been experimentally observed in nonlocal spin valves fabricated from a variety of miscible,
moment-forming ferromagnet/nonmagnet pairings, e.g., Fe/Cu and Ni$_{80}$Fe$_{20}$/Cu~\cite{Brien14NC}. Recently, through the use of thermal annealing to promote interdiffusion, we have also
shown fine control over $\kappa$ in Fe/Cu nonlocal spin valves, and directly correlated its magnitude to the Fe/Cu interdiffusion length, $\lambda_{\rm Fe}$~\cite{Brien16PRB}. Equation~(\ref{Eq:Kappa3})
quantitatively connects $\kappa(T)$ to measurable quantities for the first time, and so provides an ideal expression with which to compare to these
experiments. In the current section, we examine the experimental magnitude and $T$ dependence of $\kappa$, while varying the annealing temperature, $T_A$, in
order to tune the extent of the interfacial Kondo region. Through this analysis, and the use of Eq.~(\ref{Eq:Kappa3}), we extract an experimental value for the Elliot-Yafet probability for Kondo spin-flip scattering
$\eta$, demonstrating good agreement between the presented theory and experimental results.

In nonlocal spin valve measurements, where ferromagnetic contacts are separated by a distance $L$, $\alpha(T)$ is typically extracted by fitting $\Delta R_{\rm NL} (L,T)$ to a one-dimensional model of nonlocal spin transport~\cite{Takahashi03PRB}. In this model $\alpha_{\rm eff}$ enters as a boundary condition which principally determines the magnitude of $\Delta R_{\rm NL}$ at fixed $L$. Provided that the Kondo
region is small compared with the mesoscopic device length ($d<L$), Kondo depolarization then appears as an interfacial effect, and is manifest as a
suppression of the measured $\alpha$ at low $T$, as quantified by $\kappa$ [Eq.~(\ref{Eq:Kappa3})]. To account for this, we define an effective polarization, $\alpha_{\rm eff}=\alpha(d)$, i.e., the observed current
polarization in nonlocal spin valves with Kondo suppression present. This can be contrasted with the intrinsic polarization of the ferromagnet, $\alpha_{\rm FM}$. In this context,
$\kappa \approx \alpha_{\rm eff}/\alpha_{\rm FM}$, and so determining $\alpha_{\rm eff} (T)$, through $\Delta R_{\rm NL} (L)$, and $\alpha_{\rm FM} (T)$ yields an experimental
measure of $\kappa(T)$. In the following, we examine $\alpha_{\rm eff} (T)$ obtained from annealed Fe/Cu nonlocal spin valves. (Details of experimental fabrication and
measurement of these all-metallic nonlocal spin valves can be found in the original reports). We note that Fe/Cu represents an ideal choice of materials as Fe is miscible
and moment forming in Cu~\cite{Huchinson51PR}, with a readily accessible $T_K=$ 30 K~\cite{Mydosh93}.

\begin{figure}
\includegraphics[width=8.6cm]{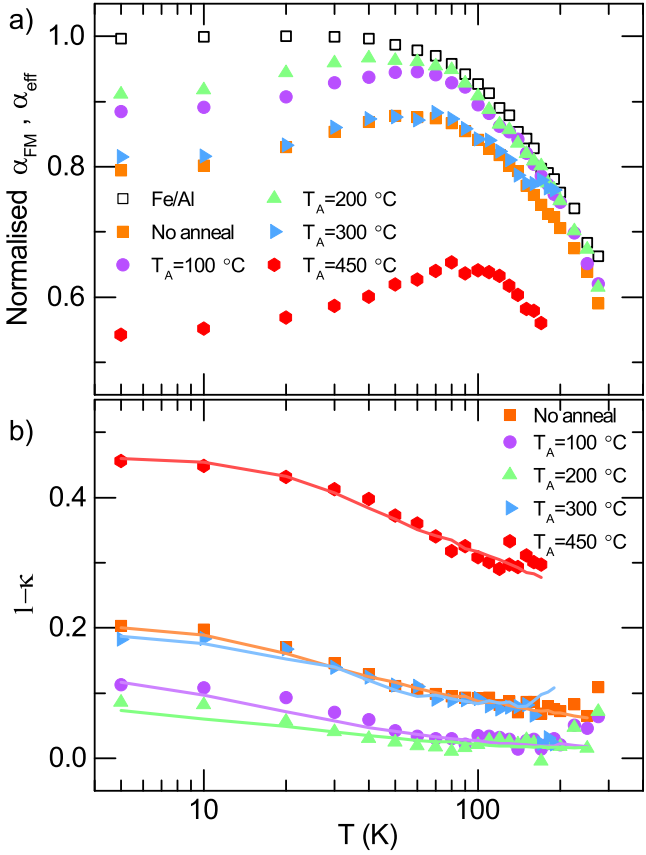}
\caption{(color online) (a) Temperature dependence of the normalized $\alpha_{\rm FM}$, obtained from Fe/Al nonlocal spin valves (open symbols), and $\alpha_{\rm eff}$ for Fe/Cu nonlocal spin valves annealed at various
temperatures (closed symbols), $T_A$. $\alpha_{\rm eff}$ is normalized to $\alpha_{\rm FM}$ using the procedure described in Appendix B. (b) Extracted $1-\kappa(T)$ from
$\alpha_{\rm eff}$. Symbols represent experimental data for various $T_A$. Solid lines in (b) are fits to the data using Eq.~(\ref{Eq:Kappa3}), with unconstrained
$T_K$ and a phenomenological Goldhaber-Gordon expression for $\rho_K$.}
\label{Fig:exptresults}
\end{figure}

To determine $\alpha_{\rm FM} (T)$ we also measured $\Delta R_{\rm NL} (L,T)$ in devices devoid of dilute impurity moments, and thus the Kondo effect. Two types of
devices were tested along these lines: nonlocal spin valves fabricated from nonmagnets that do not support local moments, e.g., Al; and nonlocal spin valves that incorporate a thin interlayer
(e.g., Al) between the ferromagnet and nonmagnet that suppresses interdiffusion and moment formation. In both types of device the normalized $\alpha_{\rm FM} (T)$ is found to be
monotonic and quantitatively similar, as shown in black squares in Fig.~\ref{Fig:exptresults}(a).

\begin{figure}
\includegraphics[width=8.6cm]{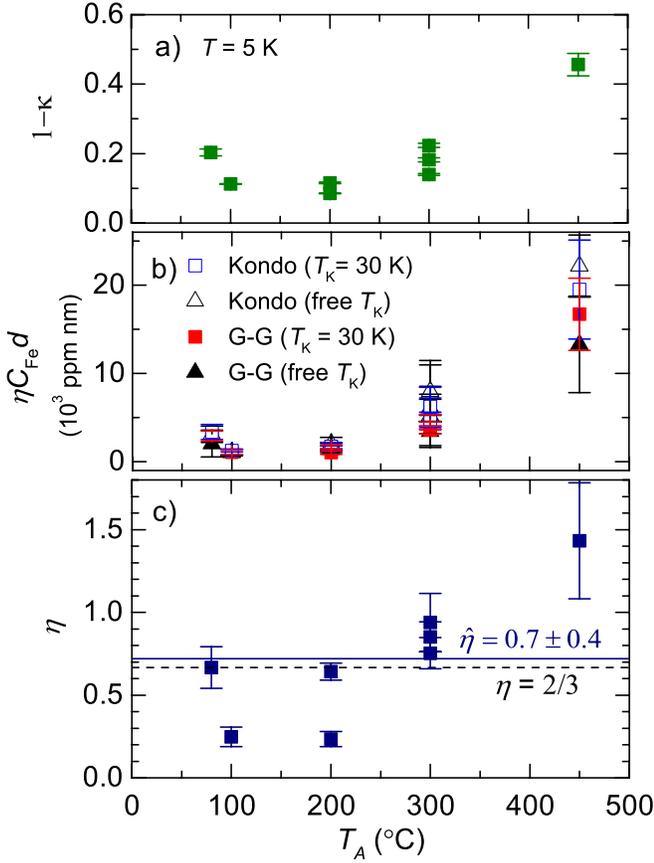}
\caption{(color online) (a) Magnitude of $1-\kappa$ for low $T$ (=5 K) as a function of annealing temperature, $T_A$. (b) Extracted values of $\eta \hat{C}_{\rm Fe} d$ using either the
Kondo (open symbols) or Goldhaber-Gordon (G-G) expressions (closed symbols) for $\rho_K$, with $T_K$ constrained (colored) or unconstrained (gray). (c) Estimated
values of $\eta$ using the results of panel b, with $\eta \hat{C}_{\rm Fe} d= \textrm{560 $\mu$mol/mol}\cdot\lambda_{\rm Fe}$ and $\lambda_{\rm Fe}$ determined from prior STEM/EDX
measurements. The error bars indicate single standard deviation uncertainties as discussed in Appendix~\ref{Appendix:error}.} \label{Fig:kappa}
\end{figure}

To avoid complications from potential interface resistance changes during annealing, as well as other inherent systematic errors between devices, we scale
$\alpha_{\rm eff}$ to $\alpha_{\rm FM} (T)$ using the method discussed in Appendix B. The resulting normalized $\alpha_{\rm eff}$ for various $T_A$ are shown in
Fig.~\ref{Fig:exptresults}(a), with the corresponding $1-\kappa(T)$, i.e. the degree of suppression, shown in Fig.~\ref{Fig:exptresults}(b). Each dataset here comes from fitting $\Delta
R_{\rm NL} (L,T)$ from devices with at least four different contact separations, ranging from $L=$ 250 nm to 5 $\mu$m. The data of Fig.~\ref{Fig:exptresults}(a) are
taken from a larger set of nonlocal spin valves (eight in total), however, for clarity we show only one curve at each $T_A$. For devices that do not support local moments $\alpha(T)$ is found to monotonically decrease with increasing $T$. In the presence of interdiffusion, however, a suppression of $\alpha(T)$ is observed at low $T$, the magnitude of which broadly increases with increasing $T_A$. Consequently, $1-\kappa(T)$ is largest at high $T_A$
and decreases with increasing measurement $T$, as would be anticipated. Figure \ref{Fig:kappa}(a) shows $1-\kappa$ at $T=$ 5 K for all samples, demonstrating
this increase in magnitude with $T_A$. Note Fig. \ref{Fig:kappa} displays data for all eight measured device batches, including multiple sets at $T_A = 200$ $^{\circ}$C and 300 $^{\circ}$C.

We now consider fitting the data of Fig.~\ref{Fig:exptresults}(b) using Eq.~(\ref{Eq:Kappa3}). In this equation $l_{N}^{\rm sf} (T)$, $\rho_N (T)$ and
$\rho_F (T)$ are measured experimentally. $\tau_N^{\rm sf}/\tau_N$ is constrained to the literature value of 950~\cite{Beuneu78PRB} --- a value we have explicitly verified in
high-conductivity nonlocal spin valve channels, where Kondo effects are negligible and phonon scattering dominates spin relaxation. We consider two expressions for $\rho_K
(T)$, the Kondo model~\cite{Kondo64PTP}~[Eq.~(\ref{Eq:rhoKK})] and the phenomenological formalism of Goldhaber-Gordon (G-G)~\cite{Goldhaber-Gordon98PRL} ~[Eq.~(\ref{Eq:rhoGG})]:
\begin{gather}
\rho_K^K= \rho_m \left(1+2N_0 J \ln \frac{T}{T^*} \right), \label{Eq:rhoKK}\\
\rho_K^{\operatorname{G-G}}=\rho_m \left[1+2N_0  J \left(  \frac{{T_K^\prime}^2}{T^2+{T_K^\prime}^2} \right)^s\ln \frac{U}{k_B T^*}  \right], \label{Eq:rhoGG}
\end{gather}
where $T_K'=T_K/\sqrt{2^{1/s}-1}$, $\rho_m=\hat{C}_{\rm Fe} 2\pi \mu_N^0 S(S+1) {N_0}^2 J^2m/3\hbar n_N^0 e^2 $ is the classical resistivity without taking account of dilute magnetic impurities, $U$ is the on-site Coulomb energy, $N_0$ is the density of states of each spin band, $J$ is the (negative) exchange parameter between the conduction electron and the magnetic impurities, and $s$ is the so-called G-G exponent which is typically taken to be 0.22~\cite{Goldhaber-Gordon98PRL}. In the expression for $\rho_m$, $\mu_N^0$ is the Fermi level of the normal metal, $\hat{C}_{\rm Fe}$ is the average impurity concentration in Fe, $S$ is the spin angular momentum of the magnetic impurities, $n_N^0$ is the density of electrons, $m$ is the effective electron mass. Here we have adapted the generalized G-G model to give agreement with Kondo's original theory. $T^*$ depends on the
limits of the energy integration of Eq.~(\ref{Eq:gamma}) and is typically taken to be either $T_K$, $U$ or $J$, depending on the theoretical treatment. In our
case, the requirement that both models be equivalent at $T=T_K$ gives $T^*=k_B {T_K}^2/U$. The Fermi energy of Cu, $\mu_N^0=$ 7 eV~\cite{Ashcroft76}, is well known, as is
$T_K=$ 30 K from $\rho_K (T)$ and susceptibility measurements~\cite{Mydosh93}. Furthermore, $J=$ 0.91 eV for Fe/Cu has been experimentally measured via field-dependent
magneto-resistance and magnetometry measurements~\cite{Monod67PRL}. From our own measurements of $\rho_K (T)$ in heavily ‘doped’ Fe/Cu nanowires, we can establish $U\approx$ 0.86
meV (10 K). Noting $\tau_N^{\rm sf}/\tau_N \gg 1$, this leaves only the product $\eta \hat{C}_{\rm Fe} d$, i.e., the weighted total number of impurities in the
Kondo region (per cross sectional area), as an unknown. Equation~(\ref{Eq:tauNsf relation}) is known to be valid only over a limited $T$ range about $T_K$, evolving to the value
dictated by the classical scattering rate at $T\gg T_K$ and the unitary limit at $T \ll T_K$. Consequently, when using this model we restrict the fit $T$
range to only consider the transition region within the data, between approximately 10 K and 100 K. In addition to considering both of these models, we also
compare to the cases where $T_K$ is allowed to be an unconstrained fitting parameter. The extracted $\eta \hat{C}_{\rm Fe} d$ from each model (through a least
mean square minimization of residuals) is shown in Fig \ref{Fig:kappa}(b). We note that the individual parameters $\eta$, $\hat{C}_{\rm Fe}$ and $d$ remain
otherwise inseparable. The solid lines in Figure~\ref{Fig:exptresults}(b) show fits of $1-\kappa(T)$ using the G-G model with unconstrained $T_K$. In general, the
overall magnitude and $T$ dependence is well captured for all $T_A$. When $T_K$ is unconstrained we find $T_K= (44\pm36)$ K for the G-G model and $T_K= (22\pm9)$ K for the Kondo model, in good agreement with the literature value of $T_K=$ 30 K. (All uncertainties in this paper are single standard deviations, the determination of which is discussed in Appendix~\ref{Appendix:error}.) The deviation in these values reflects the logarithmic dependence
of $\rho_K$ on $T$, and so the difficulty in determining $T_K$, which is often a challenge in dilute-moment metallic systems. Despite the limited range of
applicability, both expressions (i.e., for $\rho_K^K$ and $\rho_K^{\operatorname{G-G}}$) give consistent results for $\eta \hat{C}_{\rm Fe} d$, reflecting the fact that these
parameters only act to influence the \textit{magnitude} of $\kappa$, and are thus relatively insensitive to the precise $T$ dependence of the data. Examining
Fig.~\ref{Fig:kappa}(b) we see that the total number of impurities increases on annealing, with dramatic changes occurring above $T_A\approx$ 300 $^{\circ}$C, in
good agreement with our previous observations of the $T_A$ dependence of $\lambda_{\rm Fe}$ in this system~\cite{Brien16PRB}.

To place these values of $\eta \hat{C}_{\rm Fe} d$ in context, and to extract a value of $\eta$, we consider our previous work on interdiffusion in
Fe/Cu nonlocal spin valves. Fe has finite solubility in Cu, with a limit of 2600 $\mu$mol/mol at room temperature~\cite{Huchinson51PR} (based on the bulk equilibrium phase diagram), beyond which
precipitation occurs, leading to phase-segregated clusters. In the following analysis we therefore assume that regions with $\hat{C}_{\rm Fe}>$ 2600 $\mu$mol/mol do not
contain isolated dilute moments, and so do not contribute to the Kondo effect. Through energy-dispersive x-ray spectroscopy (EDX) measurements in
cross-sectional scanning transmission electron microscopy (STEM) we have previously shown that the interdiffusion profile in annealed Fe/Cu nonlocal spin valves follows
$C_{\rm Fe}(x)\propto[1-\operatorname{erf}⁡(x/\lambda_{\rm Fe}) ]/2$, and have quantitatively determined $\lambda_{\rm Fe} (T_A )$ for our devices~\cite{Brien16PRB}. Using this expression for $C_{\rm Fe}$,
considering only the dilute Kondo region below $C_{\rm Fe}<$ 2600 $\mu$mol/mol in nonlocal spin valves, and assuming $L\gg \lambda_{\rm Fe}$, yields a total number of impurity
atoms~\footnote{$\int_{x\left( C_{\rm Fe}=\textrm{2600 $\mu$mol/mol}\right)}^{\infty} [ 1-\operatorname{erf}⁡(x/\lambda_{\rm Fe}) ] dx=\textrm{560
$\mu$mol/mol}\cdot\lambda_{\rm Fe}$} (per unit cross sectional area) in the Kondo region of $\hat{C}_{\rm Fe} d= \textrm{560 $\mu$mol/mol}\cdot \lambda_{\rm Fe}$. Using this result and our
previous measurements of $\lambda_{\rm Fe}$, we are therefore able to estimate $\eta$. Duly extracted values of $\eta$ are shown in Fig.~\ref{Fig:kappa}(c) as a
function of $T_A$, with the average $\hat{\eta}=$ 0.7$\pm$0.4 indicated by the solid horizontal line. We anticipate that the simplicity of the model in accounting for the precise dispersion of Fe at the Fe/Cu interface, in addition to variability in the precise degree of interdiffusion at intermediate $T_A$, likely accounts for the dispersion in the extracted values of $\eta$, particularly at $T_A =$ 200 $^{\circ}$C. This is also likely to be the cause of the unphysical value of $\eta>1$ found at $T_A=450^{\circ}\rm{C}$ (note also the large random error). Despite the variation, we find good overall
consistency between the experimentally determined values and $\eta=$2/3 (dashed horizontal line), as calculated in the original work of Kondo~\cite{Kondo64PTP}. Given the rather simplistic assumptions made, as well as the use of four independent experimental measurements in order to determine $\hat{\eta}$, this result indicates good consistency between the model and experiment.

\section{Derivations\label{Sec:Theory}}

In this section we present mathematical derivations of the core results in
Sec.~\ref{Sec:Result-A}. First we start from the Boltzmann equation to derive the spin drift-diffusion equation at finite temperature. We take the
approach developed by Valet and Fert~\cite{Valet93PRB}. Readers not familiar with details of the Valet-Fert theory may refer to
Appendix~\ref{Sec-A:Detail-Legendre}.

We start from the Boltzmann equation for a translation-invariant system in
the two directions perpendicular to $z$. The distribution functions $f_\pm$ are a function of $z$ and $\vec{k}$. At the equilibrium, they are
$[f_s^0(z,k)]^{-1}=e^{\beta[\epsilon(\vec{k})-\mu_s^0(z)]}+1$ where
$\epsilon(\vec{k})=\hbar^2k^2/2m$ is the energy eigenvalue, $\vec{k}$ is the
crystal momentum, $k=|\vec{k}|$, $\mu_s^0(z)$ is the chemical potential at
the equilibrium, $m$ is the effective mass, $\beta=1/k_BT$ is the inverse
temperature, and $k_B$ is the Boltzmann constant. When an electric field is
applied along the $\vhat{z}$ direction, the distribution function has a small
correction to its equilibrium function
$f_s(z,\vec{k})=f_s^0(\vec{k})+g_s(z,\vec{k})$ where $g_s(\vec{k})$ is the
small correction proportional to the electric field and $s=\pm$. The
linearized Boltzmann equation at steady state under the relaxation time
approximation is
\begin{align}
&\frac{\hbar k_z}{m}\partial_zg_s(z,\vec{k})-\frac{eE}{\hbar}\partial_{k_z}f_s^0(k)\nonumber\\
&=-\frac{g_s(z,\vec{k})-g_s^{(0)}(z,k)}{\tau_s^{\rm sc}(z,k)}-\frac{g_s(z,\vec{k})-g_{-s}^{(0)}(z,k)}{\tau_s^{\rm sf}(z,k)},\label{Eq:Boltzmann}
\end{align}
where $\tau_s^{\rm sc}$ and $\tau_s^{\rm sf}$ are the relaxation times.
$\tau_s^{\rm sc}$ corresponds to spin-conserving processes $s\to s$, and
$\tau_s^{\rm sf}$ corresponds to spin-flipping processes that gives the spin
diffusion length. The sum of the spin-conserving and the spin-flipping rates
gives the electrical conductivity described below. These relaxation times are
assumed to be isotropic, that is, they only depend on the magnitude of $k$.
In general $\tau_{s}^{\rm sf}$ is independent of $s$ at the Fermi surface.
However, since the Fermi wave vector depends on $s$,
$\tau_{s}^{\rm sf}$ written \textit{as a function of $k$} is $s$ dependent.
This is also true for the Kondo contribution~\cite{Kondo64PTP}. The
relaxation times $\tau_N$ and $\tau_N^{\rm sf}$ that appear in
Sec.~\ref{Sec:Result} will be connected to these functions. $g_s^{(0)}(z,k)$
is the angle-averaged $g_s(z,\vec{k})$ over the Fermi surface with the
constant magnitude $|\vec{k}|=k$, that is,
$g_s^{(0)}(z,k)=(1/4\pi)\int_{\vec{k}=k} d\Omega_\vec{k} g_s(z,\vec{k})$
where $\Omega_\vec{k}$ is the solid angle of $\vec{k}$.

We make the approximation that the system is rotationally symmetric around the $z$ axis. In
this regime, $g(z,\vec{k})$ can be expanded by the Legendre
polynomials as $g_s(z,\vec{k})=\sum_{n=0}^\infty
g_s^{(n)}(z,k)P_n(\vhat{k}\cdot\vhat{z})$, where $P_n$ is the Legendre
polynomial. Since the Legendre polynomials form an orthonormal set of
polynomials, each coefficient satisfies the equation. Taking $n=0,1$
coefficients and neglecting the higher order contributions~\cite{Valet93PRB},
Eq.~(\ref{Eq:Boltzmann}) is equivalent to
\begin{align}
\frac{\hbar k}{3m}\partial_z g_s^{(1)}(z,k)&=-\frac{\tilde{g}_s^{(0)}(z,k)-\tilde{g}_{-s}^{(0)}(z,k)}{\tau_s^{\rm sf}(z,k)},\label{Eq:DD-0}\\
\frac{\hbar k}{m}\partial_z \tilde{g}_s^{(0)}(z,k)&=-\frac{g_s^{(1)}(z,k)}{\tau_s(z,k)},\label{Eq:DD-1}
\end{align}
where $\tilde{g}_s^{(0)}=g_s^{(0)}-eEz\partial_\epsilon f_s^0$,
$1/\tau_s=1/\tau_s^{\rm sc}+1/\tau_s^{\rm sf}$ is the total scattering-out
rate of a spin $s$ state.

Equations~(\ref{Eq:DD-0}) and (\ref{Eq:DD-1}) are the spin drift-diffusion
equations that hold at $T=0$. At $T=0$, $\tilde{g}_s^{(0)}$ and $g_s^{(1)}$
evaluated at the Fermi surface are respectively assigned to the chemical
potential and the current, with proper prefactors~\cite{Valet93PRB}. Therefore, Eqs.~(\ref{Eq:DD-0}) and (\ref{Eq:DD-1}) provide closed solutions for these
physical quantities. However, this association is not exact at finite temperature. For $T>0$, a physical quantity is not given by a value at the Fermi surface,
but is given after integrating over $k$, considering $T$ dependence of $f_s^0(k)$.

Although the temperature dependence of $f_s^0$ is very complicated, the
Sommerfeld expansion formula in Appendix~\ref{Sec-A:Sommerfeld} allows
substantial simplication. The Sommerfeld expansion formula is an
expression for integrals including the Fermi-Dirac distribution at low
temperature. In the Kondo regime, $k_BT\ll\mu_s^0$ satisfies the criterion.
Neglecting $\mathcal{O}(T^2)$, the spin density $n_s$ for each $s$ band (or
the spin chemical potential with an additional factor as shown below) is
given by
\begin{align}
n_s(z)&=\frac{e}{V}\sum_\vec{k}f_s(z,\vec{k})
\nonumber\\&=\frac{e}{2\pi^2}\int dk k^2 \tilde{g}_s^{(0)}(z,k)\nonumber\\
&\quad+\frac{m\sqrt{m\mu_s^0(z)}e^2Ez}{\sqrt{2}\pi^2\hbar^3}+\frac{m\sqrt{2m\mu_s^0(z)}\mu_s^0(z)e}{3\pi^2\hbar^3}.\label{Eq:spin acc}
\end{align}
We use here the Sommerfeld expansion formula Eq.~(\ref{Eq(C-2):Sommerfeld1})
for low temperature $k_BT\ll\mu_s^0$. 
Similarly, the current density $j_s(z)$ for $s$ band is
\begin{align}
j_s(z)&=-\frac{e}{V}\sum_\vec{k}\frac{\hbar k_z}{m} f_s(z,\vec{k})
=-\frac{e\hbar}{6\pi^2m}\int dk k^3 g_s^{(1)}(z,k).\label{Eq:currents}
\end{align}
Assuming that $\mu_s^0(z)$ is constant in space in each region (sudden
changes at the boundaries can be taken into account by matching boundary
conditions), integration of Eqs.~(\ref{Eq:DD-0}) and (\ref{Eq:DD-1}) with
the weighting factors $k^2$ and $k$ gives respectively
\begin{align}
\partial_zj_s(z)&=\frac{e}{2\pi^2}\int dk k^2\frac{\tilde{g}_s^{(0)}(z,k)-\tilde{g}_{-s}^{(0)}(z,k)}{\tau_s^{\rm sf}(z,k)},\label{Eq:DD-0 2}\\
\partial_zn_s(z)&=-\frac{em}{2\pi^2\hbar}\int dk k\frac{g_s^{(1)}(z,k)}{\tau_s(z,k)}.\label{Eq:DD-1 2}
\end{align}
Equations~(\ref{Eq:DD-0 2}) and (\ref{Eq:DD-1 2}) provide a generalized
drift-diffusion equation at low temperature.

We first briefly show that Eqs.~(\ref{Eq:DD-0 2}) and (\ref{Eq:DD-1 2}) become the conventional spin drift-diffusion equations without the Kondo
effect. That is, Valet-Fert theory holds up to $\mathcal{O}(T)$, if
there are no magnetic impurities. Without an electric field,
$\tilde{g}_s^{(0)}$ and $g_s^{(1)}$ are zero. This observation implies that
$\tilde{g}_s^{(0)}$ and $g_s^{(1)}$ are proportional to $\partial_\epsilon
f_s^0$. By the Sommerfeld expansion formula, $\partial_\epsilon f_s^0$ can be replaced by $-\delta(\mu_s^0-\epsilon)$, neglecting $\mathcal{O}(T^2)$. Thus,
the integrations Eqs.~(\ref{Eq:DD-0 2}) and (\ref{Eq:DD-1 2}) are nothing but evaluations at the Fermi surface. Therefore, Eqs.~(\ref{Eq:DD-0 2}) and
(\ref{Eq:DD-1 2}) are equivalent to Eqs.~(\ref{Eq:DD-0}) and (\ref{Eq:DD-1})
up to $\mathcal{O}(T)$.

Now we connect the quantities that appear in Eqs.~(\ref{Eq:DD-0}) and
(\ref{Eq:DD-1}) to physical quantities. First we define the electrical
conductivity $\sigma_s(z)$ and the spin diffusion length $l_s(z)$ for each
spin band $s$, which are respectively given by
\begin{align}
\sigma_s(z)&=\frac{e^2(k_s^F)^3\tau_s(z,k_s^F)}{6\pi^2m},\label{Eq:sigma to tau}\\
{l_s}^2(z)&=\frac{\hbar^2 (k_s^F)^2}{3m^2}\tau_s(z,k_s^F)\tau_s^{\rm sf}(z,k_s^F),\label{Eq:l to tau}
\end{align}
where $k_s^F=\sqrt{2m\mu_s^0}/\hbar^2$ is the Fermi wave vector. Here the
electrical conductivity is equivalent to the Drude conductivity
$\sigma_s=n_s^0 e^2 \tau_s(k_s^F)/m$ where $n_s^0= (k_s^F)^3/6\pi^2$ is $n_s$
without an electric field. Similarly, the spin diffusion length is related to
the diffusion constant by ${l_s}^2=D_s\tau_s^{\rm sf}(k_s^F)$ where
$D_s=\hbar^2 (k_s^F)^2\tau_s(z,k_s^F)/3m^2$. Next, we define the
electrochemical potential $\mu_s=(2\pi^2\hbar^2/emk_s^F)n_s$. The factor
arises from the ratio between $\int d\epsilon$ and $(e/V)\sum_\vec{k}$ in
Eq.~(\ref{Eq:spin acc}). With these definitions, Eqs.~(\ref{Eq:DD-0}) and
(\ref{Eq:DD-1}) become equivalent to Eqs.~(\ref{Eq:V-F eq1}) and (\ref{Eq:V-F eq2}).

The situation changes in the presence of dilute
magnetic impurities. We show that Eqs.~(\ref{Eq:V-F eq1}) and
(\ref{Eq:V-F eq2}) are still valid after replacement of Eqs.~(\ref{Eq:renormalization sigma}) and (\ref{Eq:renormalization l}) [or
equivalently Eqs.~(\ref{Eq:tauNF relation}) and (\ref{Eq:tauNsf relation})]
and give explicit expressions for $\tau_K^{\rm eff}$ in particular regimes. Since the
Kondo effect occurs in the normal metal, we use the subscript $N$ but drop
the spin-dependent subscript $s$. That is, $\tau_N(z,k)$ and $\tau_N^{\rm
sf}(z,k)$ are the relaxation times in the normal metal, which are
spin independent. $\tau_N$ and $\tau_N^{\rm sf}$ that appear in
Sec.~\ref{Sec:Result} are those evaluated at the Fermi level $k=k_N^F$. In
the presence of dilute magnetic impurities, additional relaxation rates arise
due to the impurities. We denote these by the subscript $K$. In the Boltzmann
equation Eq.~(\ref{Eq:Boltzmann}), the relaxation times change by
\begin{align}
\frac{1}{\tilde{\tau}_N^{\rm sc}(z,k)}&=\frac{1}{\tau_N^{\rm sc}(z,k)}+\frac{1}{\tau_K^{\rm sc}(z,k)},\\
\frac{1}{\tilde{\tau}_N^{\rm sf}(z,k)}&=\frac{1}{\tau_N^{\rm sf}(z,k)}+\frac{1}{\tau_K^{\rm sf}(z,k)},
\end{align}
where $\tilde{\tau}_N^{\rm sc}$ and $\tilde{\tau}_N^{\rm sf}$ are the
modified relaxation times due to the Kondo effect. Kondo~\cite{Kondo64PTP}
computed explicitly the total scattering rate change
${\tau_K}^{-1}=(\tau_K^{\rm sc})^{-1}+(\tau_K^{\rm sf})^{-1}$ given by
\begin{align}
\frac{1}{\tau_K(z,k)}&=\frac{2\pi\mu_N^0S(S+1)\hat{C}N_0^2J^2}{3\hbar}[1+2J\gamma(\epsilon)],\label{Eq:tk}\\
\gamma(\epsilon)&=\frac{1}{V}\sum_\vec{k}\frac{f_N^0(\vec{k})}{\epsilon(\vec{k})-\epsilon},\label{Eq:gamma}
\end{align}
where $S$ is the spin angular momentum of the magnetic impurities,
$N_0=mk_N^F/2\pi^2\hbar^2$ is the density of states of each spin band, $J$ is
the (negative) exchange parameter between the conduction electron and the
magnetic impurities, and $\hat{C}$ is the average impurity concentration, which is the
density of impurities divided by the density of electrons
$2n_N^0=(k_N^F)^3/3\pi^2$. The units of $J$ are $\mathrm{J\cdot m^3}$ to make
$N_0J$ dimensionless. To be explicit, we set the Kondo Hamiltonian to be
$H_K=-(4J/V)\sum_{\vec{k}'\vec{k}}(\Psi_{\vec{k}'}^\dagger\vec{\sigma}\Psi_\vec{k})\cdot(\Psi_{d}^\dagger\vec{\sigma}\Psi_{d})$,
where $\Psi_{\vec{k}}^\dagger$ and $\Psi_{d}^\dagger$ are respectively the
electron creation operator of conduction electrons with momentum $\vec{k}$
and electrons in the impurity state $d$ and $\vec{\sigma}$ is the Pauli
matrix.

Since the Kondo theory is a perturbation theory, its contribution to the
spin-flip rate $(\tau_K^{\rm sf})^{-1}$ is likely to be proportional to
Eq.~(\ref{Eq:tk}) as Kondo showed~\cite{Kondo64PTP}. We introduce $\eta$, the spin-flip probability during each Kondo scattering event by
\begin{equation}
\frac{1}{\tau_K^{\rm sf}(z,k)}=\frac{\eta}{\tau_K(z,k)}.
\end{equation}
The value of $\eta$ is determined by the Fermi surface geometry. For a
spherical Fermi surface that we use here, Kondo~\cite{Kondo64PTP} showed that
$(\tau_K^{\rm sf})^{-1}$ is twice $(\tau_K^{\rm sc})^{-1}$~\footnote{See
Eq.~(12) of \ocite{Kondo64PTP}.}. This indicates that $\eta=2/3$ for this
case.

The low temperature behavior of $\tau_K(z,k)$ requires careful treatment. Since
$\gamma(\epsilon)$ diverges at the Fermi level, na\"{i}ve application of the
Sommerfeld expansion gives divergences. Although the integrals in
Eqs.~(\ref{Eq:DD-0 2}) and (\ref{Eq:DD-1 2}) seem surprisingly difficult to
perform without the Sommerfeld expansion, a low temperature approximation
allows it. In
Appendix~\ref{Sec-A:Kondo integrals}, we slightly generalize Kondo's
approach to extract the logarithmic dependence of the Kondo resistivity to
show that the following replacement is valid under the integration over the
energy.
\begin{equation}
\gamma(\epsilon)\partial_\epsilon f_N^0\to N_0\ln\frac{k_BT}{\mu_N^0}[-\delta(\epsilon-\mu_N^0)],\label{Eq:Kondo integral}
\end{equation}
giving rise to a $\ln T$ contribution. With this rule, Eqs.~(\ref{Eq:V-F eq1}) and (\ref{Eq:V-F eq2}) still hold under the following replacement.
\begin{align}
\frac{1}{\tau_N(z,k_s^F)}&\to\frac{1}{\tilde{\tau}_N(z,k_s^F)}\equiv\frac{1}{\tau_N(z,k_s^F)}+\frac{1}{\tau_K^{\rm eff}},\\
\frac{1}{\tau_N^{\rm sf}(z,k_s^F)}&\to\frac{1}{\tilde{\tau}_N^{\rm sf}(z,k_s^F)}\equiv\frac{1}{\tau_N^{\rm sf}(z,k_s^F)}+\frac{\eta}{\tau_K^{\rm eff}},
\end{align}
which are nothing but Eqs.~(\ref{Eq:tauNF relation}) and (\ref{Eq:tauNsf
relation}). Here
\begin{equation}
\frac{1}{\tau_K^{\rm eff}}=\frac{2\pi\mu_N^0S(S+1)\hat{C}{N_0}^2J^2}{3\hbar}\left(1+2N_0J\ln\frac{k_BT}{\mu_N^0}\right).\label{Eq:Kondo rate}
\end{equation}

\section{Summary\label{Sec:Summary}}

In order to take account for the Kondo effect in spin transport, we derive a modified drift-diffusion equation from the Boltzmann equation explicitly allowing for finite temperature. The complicated finite temperature theory is projected to a low temperature regime (compared to the Fermi temperature). We show that the Valet-Fert drift-diffusion equation holds both at finite $T$ and in the presence of spin scattering from dilute magnetic impurities, once the electrical conductivity and spin diffusion length are renormalized as functions of temperature. This represents a useful result; as a consequence, dilute magnetic impurity scattering beyond the semiclassical limit can indeed be described in a simple Elliot-Yafet-like form with a direct proportionality between $\tau^{\rm sf}_K$ and $\tau_K$, as originally indicated by Kondo. The modified drift-diffusion equation has a remarkably compact form given the complexity of the higher order many-body interactions involved.

By solving the drift-diffusion equation for an illustrative regime, we show additional spin relaxation in the presence of the Kondo effect at a ferromagnet/nonmagnet interface. Kondo scattering is found to be highly efficient at spin relaxation, due to the high probability of spin flip $(\eta=2/3)$ compared with other scattering mechanisms (c.f. $\eta_{\rm phonon}\approx 1/1000$). Since the spin-flip rate is much lower than the momentum scattering rate in the absence of the Kondo effect, such a high probability caused by the Kondo effect can significantly reduce the spin diffusion length, even when there is negligible change to the conductivity. This is confirmed experimentally by the large value of $\eta\approx0.7$ observed, in good agreement with Kondo's original work. We hope this, in addition to the explicit derivation of Eq.~(\ref{Eq:renormalization l}), further validates the semiclassical model of Ref.~\cite{Batley15PRB} in also determining the Kondo contributions to $l_N$.  Note again that the fitting procedure used here relies on four independent quantities that are experimentally extracted, as well as approximations regarding the precise distribution of magnetic moments within the Kondo region. Examining Fig.~\ref{Fig:kappa}, one can see a weak dependence of $\eta$ on $T_A$. This is very likely due to such simplifications. Indeed, the possibilities of Fe segregation and cluster formation on annealing, dilute impurity migration to grain boundaries, and inter-moment correlations at high concentrations, as well as examining the precise phase equilibrium beyond the thermodynamic limit are entirely overlooked, and could greatly complicate the situation. Nevertheless, agreement between the simple model and experiment is highly satisfactory.

One observation worth mentioning is the form of Eq.~(\ref{Eq:Kappa3}), particularly the fact the signal suppression is linear in both $\hat{C}_{\rm Fe}$ \textit{and} $d$. This clarifies one of the fundamental difficulties previously experienced within the field. That is, determining the precise location of the anomalous relaxation mechanism. Previous reports have stated relaxation occurring at the ferromagnet/nonmagnet interface (as we discuss here), throughout the channel~\cite{Villamor13PRB,Batley15PRB}, or at surfaces~\cite{Kimura08PRL,Zou12APL,Idzuchi12APL}, with similar magnitudes of Kondo suppression in each case. To first order it is the product $\hat{C}_{\rm Fe}d$ (total number of impurities per cross-sectional area) that determines suppression, and so similar magnitudes may be observed \textit{either} due to a high-impurity-concentration narrow region (e.g. an interfacial effect), or an extended low concentration region (i.e., low doping levels throughout the channel itself). For the case where magnetic impurities extend throughout the channel, i.e. in the limit where $d\geq \tilde{l}_N$, the approximations made in obtaining Eq.~(\ref{Eq:Kappa3}) will no longer be appropriate. Instead separation-dependent measurements of $\Delta R_{\rm{NL}}$ on mesoscopic lengthscales (i.e. comparable to $\tilde{l}_N$) will follow the standard non-local spin transport equations, now with the modified value of $l_N$ given by Eq.~\ref{Eq:renormalization l}. For low impurity levels this results in comparable magnitudes of suppression to those seen here. Rather than serendipitous, it is entirely expected that both the interfacial effects discussed here and `contaminated' channel devices observe similar signal contributions from Kondo effects. This highlights the care that must be taken when fitting $\Delta R_{\rm NL}(L,T)$ to resolve the contributions from interfacial (manifest in the extracted $\alpha$) and bulk (manifest in $l_N$) Kondo relaxation, in the likely scenario where both cause $\Delta R_{\rm NL}$ to be suppressed by comparable amounts.

Having now determined the theoretical and experimental $T$-dependence of Kondo spin scattering in nonlocal spin valves, this opens the path to using the Kondo effect to better understand magnetic and nonmagnetic impurity spin relaxation. In particular, the clearly identifiable $T$-dependence may now be used as a signature to quantitatively determine the contribution of dilute moments to relaxation in all-metal systems.

\begin{acknowledgments}
The authors acknowledge P. Haney and O. Gomonay for critical reading of the manuscript. K.W.K. was supported by the Cooperative Research Agreement between the University of Maryland and the National Institute of Standards and Technology, Center for Nanoscale Science and Technology (70NANB10H193), through the University of Maryland. K.W.K also acknowledges support by Basic Science Research Program through the National Research Foundation of Korea (NRF) funded by the Ministry of Education (2016R1A6A3A03008831), Alexander von Humboldt Foundation, the ERC Synergy Grant SC2 (No. 610115), and the Transregional Collaborative Research Center (SFB/TRR) 173 SPIN+X. Work at the University of Minnesota was funded by Seagate Technology Inc., the University of Minnesota (UMN) NSF MRSEC under award DMR-1420013, and DMR-1507048. L.O'B. acknowledges a Marie Curie International Outgoing Fellowship within the 7th European Community Framework Programme (project No. 299376).
\end{acknowledgments}

\begin{appendix}
\begin{widetext}
\section{Solution of the spin drift-diffusion equation\label{Sec-A:BC}}

In \ocite{Valet93PRB}, the general solutions for Eqs.~(\ref{Eq:V-F eq1}) and
(\ref{Eq:V-F eq2}) are given by
\begin{align}
\mu_+(z)-\mu_-(z)&=\left\{\begin{array}{cl}
              A_F e^{z/l_F^{\rm sf}}&~\mathrm{for}~z<0,\\
              \tilde{A}_N e^{z/\tilde{l}_N^{\rm sf}}+\tilde{B}_N e^{-z/\tilde{l}_N^{\rm sf}} &~\mathrm{for}~0<z<d,\\
              B_N e^{-z/l_N^{\rm sf}} &~\mathrm{for}~z>d,\\
            \end{array}\right.\\
\sigma_+(z)\mu_+(z)+\sigma_-(z)\mu_-(z)&=\left\{\begin{array}{cl}
              C_Fz+D_F &~\mathrm{for}~z<0,\\
              \tilde{C}_Nz+\tilde{D}_N&~\mathrm{for}~0<z<d,\\
              C_Nz+D_N &~\mathrm{for}~z>d.\\
            \end{array}\right.
\end{align}
In this section we determine the coefficients satisfying the transparent
boundary conditions
\begin{equation}
\mu_s(z=-0)=\mu_s(z=+0),~\mu_s(z=d-0)=\mu_s(z=d+0),~j_s(z=-0)=j_s(z=+0),~j_s(z=d-0)=j_s(z=d+0).
\end{equation}
There are eight boundary conditions (note that $s=\pm$) although there are
ten unknown coefficients. Therefore, two more conditions are required. The
first one originates from a constant shift of the chemical potential. Since
the drift-diffusion equation is invariant under a constant shift of the
chemical potential, we can put $\tilde{D}_N=0$ without any loss of
generality. The second one originates from the homogeneity of the
drift-diffusion equation. The drift-diffusion equation is invariant under
multiplication by a constant factor to $\mu_s$. The applied electrical
current defined by
\begin{equation}
ej_{\rm app}=ej_+(z=-\infty)+ej_-(z=-\infty)=C_F,
\end{equation}
is an experimentally controllable quantity that fixes the multiplication
factor.

Now we apply the boundary conditions. Instead of applying the continuity of
each functions, we can apply it with their independent linear combinations.
Note that $\mu_+(z)-\mu_-(z)$ is already given above and $j_+(z)+ j_-(z)$ is
nothing but the derivative of $\sigma_+(z)\mu_+(z)+\sigma_-(z)\mu_-(z)$.
Continuity of these functions at $z=0$ and $z=d$ gives the following four
conditions.
\begin{equation}
A_F=\tilde{A}_N+\tilde{B}_N,~\tilde{A}_Ne^{d/\tilde{l}_N^{\rm sf}}+\tilde{B}_Ne^{-d/\tilde{l}_N^{\rm sf}}=B_N e^{-d/l_N^{\rm sf}},~\tilde{C}_N=C_N=ej_{\rm
app}.
\end{equation}
We now put these into the solution and obtain
\begin{align}
\mu_+(z)-\mu_-(z)&=\left\{\begin{array}{cl}
              (\tilde{A}_N+\tilde{B}_N) e^{z/l_F^{\rm sf}}&~\mathrm{for}~z<0,\\
              \tilde{A}_N e^{z/\tilde{l}_N^{\rm sf}}+\tilde{B}_N e^{-z/\tilde{l}_N^{\rm sf}} &~\mathrm{for}~0<z<d,\\
              (\tilde{A}_Ne^{d/\tilde{l}_N^{\rm sf}}+\tilde{B}_Ne^{-d/\tilde{l}_N^{\rm sf}}) e^{-(z-d)/l_N^{\rm sf}} &~\mathrm{for}~z>d,\\
            \end{array}\right.\label{Eq(A):solutions2}\\
\sigma_+(z)\mu_+(z)+\sigma_-(z)\mu_-(z)&=\left\{\begin{array}{cl}
              ej_{\rm app}z+D_F &~\mathrm{for}~z<0,\\
              ej_{\rm app}z&~\mathrm{for}~0<z<d,\\
              ej_{\rm app}z+D_N &~\mathrm{for}~z>d.\\
            \end{array}\right.\label{Eq(A):solutions1}
\end{align}

Now we apply the continuity of $\mu_+(z)+\mu_-(z)$. After some algebra,
\begin{equation}
\mu_+(z)+\mu_-(z)=\left\{\begin{array}{cl}
\displaystyle\frac{2ej_{\rm app}z+2D_F}{\sigma_{+,F}+\sigma_{-,F}}-\frac{\sigma_{+,F}-\sigma_{-,F}}{\sigma_{+,F}+\sigma_{-,F}}(\tilde{A}_N+\tilde{B}_N)
e^{z/l_F^{\rm sf}}&~\mathrm{for}~z<0,\\[10pt]
              \displaystyle\frac{ej_{\rm app}z}{\tilde{\sigma}_N} &~\mathrm{for}~0<z<d,\\[10pt]
              \displaystyle\frac{ej_{\rm app}z+D_N}{\sigma_N} &~\mathrm{for}~z>d.\\
            \end{array}\right.
\end{equation}
Continuity at $z=0$ and $z=d$ gives
\begin{equation}
D_F=\frac{\sigma_{+,F}-\sigma_{-,F}}{2}(\tilde{A}_N+\tilde{B}_N),~D_N=\left(\frac{\sigma_N}{\tilde{\sigma}_N}-1\right)ej_{\rm app}d.\label{Eq(A):D}
\end{equation}
Then $\tilde{A}_N$ and $\tilde{B}_N$ are the only remaining coefficients. We
now apply continuity of $j_+(z)-j_-(z)$. After some algebra,
\begin{equation}
\sigma_+(z)\mu_-(z)-\sigma_-(z)\mu_-(z)=\left\{\begin{array}{cl}
\displaystyle \frac{\sigma_{+,F}-\sigma_{-,F}}{\sigma_{+,F}+\sigma_{-,F}}(ej_{\rm
app}z+D_F)+\frac{2\sigma_{+,F}\sigma_{-,F}}{\sigma_{+,F}+\sigma_{-,F}}(\tilde{A}_N+\tilde{B}_N)e^{z/l_F^{\rm sf}} &~\mathrm{for}~z<0,\\[10pt]
              \displaystyle \tilde{\sigma}_N(\tilde{A}_N e^{z/\tilde{l}_N^{\rm sf}}+\tilde{B}_N e^{-z/\tilde{l}_N^{\rm sf}})&~\mathrm{for}~0<z<d,\\[10pt]
\displaystyle \sigma_N(\tilde{A}_Ne^{d/\tilde{l}_N^{\rm sf}}+\tilde{B}_Ne^{-d/\tilde{l}_N^{\rm sf}}) e^{-(z-d)/l_N^{\rm sf}}&~\mathrm{for}~z>d.\\            \end{array}\right.
\end{equation}
Continuity of the derivatives at $z=0$ and $z=d$ gives
\begin{equation}
\frac{\sigma_{+,F}-\sigma_{-,F}}{\sigma_{+,F}+\sigma_{-,F}}ej_{\rm
app}+\frac{2\sigma_{+,F}\sigma_{-,F}}{\sigma_{+,F}+\sigma_{-,F}}\frac{\tilde{A}_N+\tilde{B}_N}{l_F^{\rm sf}}=\frac{\tilde{\sigma}_N}{\tilde{l}_N^{\rm
sf}}(\tilde{A}_N -\tilde{B}_N ),
\end{equation}
\begin{equation}
\frac{\tilde{\sigma}_N}{\tilde{l}_N^{\rm sf}}(\tilde{A}_N e^{d/\tilde{l}_N^{\rm sf}}-\tilde{B}_N e^{-d/\tilde{l}_N^{\rm sf}})=-\frac{\sigma_N}{l_N^{\rm
sf}}(\tilde{A}_Ne^{d/\tilde{l}_N^{\rm sf}}+\tilde{B}_Ne^{-d/\tilde{l}_N^{\rm sf}}),
\end{equation}
the solutions of which are
\begin{align}
\tilde{A}_N&=e^{-2d/\tilde{l}_N^{\rm sf}}\frac{l_N^{\rm sf}\tilde{\sigma}_N-\tilde{l}_N^{\rm sf}\sigma_N}{l_N^{\rm sf}\tilde{\sigma}_N+\tilde{l}_N^{\rm
sf}\sigma_N}\tilde{B}_N,\\
\tilde{B}_N&=-\frac{\sigma_{+,F}-\sigma_{-,F}}{\sigma_{+,F}+\sigma_{-,F}}ej_{\rm app}\left[\left(\frac{2}{l_F^{\rm
sf}}\frac{\sigma_{+,F}\sigma_{-,F}}{\sigma_{+,F}+\sigma_{-,F}}+\frac{\tilde{\sigma}_N}{\tilde{l}_N^{\rm sf}}\right)-e^{-2d/\tilde{l}_N^{\rm
sf}}\frac{l_N^{\rm sf}\tilde{\sigma}_N-\tilde{l}_N^{\rm sf}\sigma_N}{l_N^{\rm sf}\tilde{\sigma}_N+\tilde{l}_N^{\rm sf}\sigma_N}\left(\frac{2}{l_F^{\rm
sf}}\frac{\sigma_{+,F}\sigma_{-,F}}{\sigma_{+,F}+\sigma_{-,F}}-\frac{\tilde{\sigma}_N}{\tilde{l}_N^{\rm sf}}\right)\right]^{-1}.
\end{align}
$D_F$ and $D_N$ are determined by Eq.~(\ref{Eq(A):D}). Thus, we determine all
coefficients of Eqs.~(\ref{Eq(A):solutions1}) and (\ref{Eq(A):solutions2}).
\end{widetext}

\section{Scaling procedure of the experimental data\label{Sec-A:Scaling}}

Due to inevitable sample-to-sample variations, and potential changes in interface resistance, limited information can be extracted from changes in the
absolute magnitude of $\alpha_{\rm eff}$ on annealing. This however does not preclude an analysis of the changes to Kondo depolarization, provided a method is
established to appropriately scale $\alpha_{\rm eff}(T)$. In this section we will outline the procedure applied to reach the scaled data of Fig.~\ref{Fig:exptresults}(a).

The Kondo expression of Eq.~(\ref{Eq:rhoKK}) is valid only over a narrow range about $T\approx T_K$, and one of the major successes of the G-G formalism was to
accurately describe the evolution of $\rho_K (T)$ from low ($T\ll T_K$) to high T ($T\gg T_K$). It is worth noting at this point the limiting values of
$\rho_K$ in these two regimes. At low $T$ the Kondo effect saturates towards a constant scattering rate as the unitary limit is reached, which Kondo proposed
to give $\rho_K\rightarrow \rho_m \left[ 1+2N_0 J \ln⁡ (U/T^*) \right]$. At high $T$ the effect is negligible and $\rho_K$ tends to the classical constant
expression for spin-flip scattering via exchange with the ferromagnetic impurity $\rho_K\rightarrow \rho_m$ (i.e. the Korringa rate). Although we cannot \textit{a
priori} determine the magnitude in these two regimes for our experimental data, by considering the data of reference~\cite{Kondo64PTP} and the
transition temperatures between the three regimes we can deduce $\rho_K (T\ll T_K )/\rho_K (T\gg T_K)=1+2N_0 J \ln ⁡(U/T^* )\approx$ 1.8.

Using Eq.~(\ref{Eq:Kappa3}) we may obtain an experimental estimate of $\rho_K(T)$ at each $T_A$, which is explicitly dependent upon $\alpha_{\rm eff}(T)$. (It is worth noting that $\rho_N$, $\rho_F$, $l_{F}^{sf}$, $l_{N}^{sf}$ and $\alpha_{\rm FM}$ are measured directly, while $\eta$, $d$ and $\tau_N^{sf}/\tau_N$ are all $T$-independent, leaving only the scaled value of $\alpha_{\rm eff}$ undetermined in Eq.~(\ref{Eq:Kappa3}).) To appropriately normalize $\alpha_{\rm eff} (T)$ to $\alpha_{\rm FM} (T)$ we linearly scale $\alpha_{\rm eff} (T)$ (and consequently modify $\rho_K(T)$) in order to reach the correct ratio of $\rho_K$ at low- and high-$T$ (i.e. $\rho_K (T\ll T_K )/\rho_K (T\gg T_K)=1.8$). Note, that since scaling $\alpha_{\rm eff}$ in this way only ensures the correct ratio of Kondo to classical scattering, we may still fit $1-\kappa$ to obtain the \textit{magnitude} of the scattering (both Kondo and classical) and therefore deduce $\eta$.

Once we have established correct normalization for a single dataset (in this case the unannealed data), we may normalize the remaining data by observing
the following relationship from Eq.\ref{Eq:rhoKK} and Eq.\ref{Eq:rhoGG}:
\begin{equation}
    \frac{\rho_K^i (T)d^i}{\rho_K^j (T)d^j}=\frac{d^i {\hat{C}_{\rm Fe}}^i}{d^j {\hat{C}_{\rm Fe}}^j} = \textrm{const}. \label{Eq:normal}
    \end{equation}
Where, the superscript $i,j$ denotes values for different $T_A$. Note this relationship exploits the fact that annealing only serves to increase the magnitude of $d$ and $C_{\rm Fe}$ in $\rho_K$, both of which are $T$ independent, thus the functional form of $\rho_K(T)$ is independent of $T_A$. From Eq.~(\ref{Eq:Kappa3}):
\begin{equation}
 \rho_K^i (T)d^i=(1-r^i \kappa^i) \left[ \frac{2\eta (\tau_{N}^{\rm sf}/\tau_N)}{\rho_N^i} \left( \frac{1}{\lambda_N^i}-\frac{1}{\lambda_N^i+\frac{\lambda_F
\rho_F^i}{(1-{\alpha}^2 ) \rho_N^i}} \right) \right]^{-1}.\label{Eq(B):normalization}
\end{equation}
Here $r^i$ is the scaling factor for $\alpha_{\rm eff} (T)$. Since the ratio of $\rho_K d$ is constant, we minimize
the standard deviation of expression \ref{Eq:normal} by varying $r^i$, to ensure $\alpha_{\rm eff} (T)$ is correctly scaled at each $T_A$.

\begin{widetext}
\section{Mathematical details for the derivation\label{Sec-A:Detail}}
\subsection{Legendre decomposition of the Boltzmann equation\label{Sec-A:Detail-Legendre}}

We first expand the first term in Eq.~(\ref{Eq:Boltzmann}) by the Legendre
polynomial. The Bonnet recursion formula is useful to do this.
\begin{equation}
xP_n(x)=\frac{n+1}{2n+1}P_{n+1}(x)+\frac{n}{2n+1}P_{n-1}(x),
\end{equation}
for $n\ge1$.
\begin{align}
\frac{\hbar k_z}{m}\partial_zg_s(z,\vec{k})&=\frac{\hbar k}{m}\sum_{n=0}^\infty \partial_zg_s^{(n)}(z,k)\cos\theta_\vec{k}P_n(\cos\theta_\vec{k})\nonumber\\
&=\frac{\hbar k}{m}\partial_zg_s^{(0)}(z,k)P_1(\cos\theta_\vec{k})+\frac{\hbar k}{m}\sum_{n=1}^\infty
\partial_zg_s^{(n)}(z,k)\left[\frac{n+1}{2n+1}P_{n+1}(\cos\theta_\vec{k})+\frac{n}{2n+1}P_{n-1}(\cos\theta_\vec{k})\right]\nonumber\\
&=\frac{\hbar k}{m}\sum_{n=0}^\infty
\left[\frac{n}{2n-1}\partial_zg_s^{(n-1)}(z,k)+\frac{n+1}{2n+3}\partial_zg_s^{(n+1)}(z,k)\right]P_{n}(\cos\theta_\vec{k}).
\end{align}

The second term in Eq.~(\ref{Eq:Boltzmann}) is
\begin{equation}
-\frac{eE}{\hbar}\partial_{k_z}f_s^0(k)=-\frac{eE}{\hbar}\partial_kf_s^0(k)\times P_1(\cos\theta_\vec{k}).
\end{equation}

The right-hand side of Eq.~(\ref{Eq:Boltzmann}) is
\begin{equation}
-\frac{g_s(z,\vec{k})-\overline{g_s(z,k)}}{\tau_s^{\rm sc}(z,k)}-\frac{g_s(\vec{k})-\overline{g_{-s}(z,k)}}{\tau^{\rm
sf}(z,k)}=-\frac{g_s^{(0)}(z,k)-g_{-s}^{(0)}(z,k)}{\tau^{\rm sf}(z,k)}-\frac{1}{\tau_s(z,k)}\sum_{n=1}^\infty g_s^{(n)}P_n(\cos\theta_\vec{k}),
\end{equation}
where $1/\tau_s=1/\tau_s^{\rm sc}+1/\tau^{\rm sf}$.

In summary, Eq.~(\ref{Eq:Boltzmann}) is equivalent to
\begin{align}
&\sum_{n=0}^\infty\left[\frac{\hbar k}{m}\frac{n}{2n-1}\partial_zg_s^{(n-1)}(z,k)+\frac{\hbar
k}{m}\frac{n+1}{2n+3}\partial_zg_s^{(n+1)}(z,k)-\frac{eE}{\hbar}\partial_kf_s^0(k)\delta_{n,1}\right]P_n(\cos\theta_\vec{k})\nonumber\\
&=-\frac{g_s^{(0)}(z,k)-g_{-s}^{(0)}(z,k)}{\tau^{\rm sf}(z,k)}-\frac{1}{\tau_s(z,k)}\sum_{n=1}^\infty g_s^{(n)}P_n(\cos\theta_\vec{k}).
\end{align}
Since $\{P_n\}$ forms an orthogonal set of polynomials, each coefficient
should satisfy the equation. In \ocite{Valet93PRB}, if the spin diffusion
length is much larger than the mean free path of conduction electrons,
$g_s^{(2)}$ (and higher order terms) can be neglected. The coefficients of
$P_0$ and $P_1$ gives Eqs.~(\ref{Eq:DD-0}) and (\ref{Eq:DD-1}), once
$g_s^{(2)}$ is neglected.

\subsection{Sommerfeld expansion formula\label{Sec-A:Sommerfeld}}

In this section, we present the Sommerfeld formula for low temperature. In
the main text, we keep terms up to $\mathcal{O}(T)$, we here present the
formula up to $\mathcal{O}(T^3)$ for more motivated readers.

The Sommerfeld expansion formula for a differentiable function $H$ is
\begin{equation}
\int \frac{H(\epsilon)}{e^{\beta(\epsilon-\mu)}+1} d\epsilon=\int^\mu H(\epsilon)d\epsilon+\frac{\pi^2}{6\beta^2}H'(\mu)+\mathcal{O}(T^4).
\end{equation}
In a compact form, as far as quantities after integration over $\epsilon$ is
concerned,
\begin{equation}
\frac{1}{e^{\beta(\epsilon-\mu)}+1}=\Theta(\mu-\epsilon)+\frac{\pi^2}{6\beta^2}\delta'(\mu-\epsilon)+\mathcal{O}(T^4).\label{Eq(C-2):Sommerfeld1}
\end{equation}
When a transport property is concerned, it is convenient to take the
derivative with respect to $\epsilon$.
\begin{equation}
\frac{\partial}{\partial\epsilon}\frac{1}{e^{\beta(\epsilon-\mu)}+1}=-\delta(\mu-\epsilon)-\frac{\pi^2}{6\beta^2}\delta''(\mu-\epsilon)+\mathcal{O}(T^4).\label{Eq(C-2):Sommerfeld2}
\end{equation}

\subsection{Integrals including the Kondo scattering rate\label{Sec-A:Kondo integrals}}
In this section, we perform the following integration for a general
$G(\epsilon)$
\begin{equation}
\int d\epsilon G(\epsilon)\gamma(\epsilon)\partial_\epsilon f^0,\label{Eq(C-3):Kondo integral}
\end{equation}
where $f^0=[1+e^{\beta(\epsilon-\mu^0)}]^{-1}$ and $\gamma(\epsilon)$ is defined by Eq.~(\ref{Eq:gamma}). Here and from now on, we
denote $k=\sqrt{2m\epsilon}/\hbar$, $k'=\sqrt{2m\epsilon'}/\hbar$ and so on,
appearing in integrations with respect to $\epsilon$ and $\epsilon'$. Also,
we define $k^F=\sqrt{2m\mu^0}/\hbar$ which is the Fermi wave vector. We generalize the approach taken by Kondo~\cite{Kondo64PTP} here.

First we perform the summation in $\gamma(\epsilon)$.
\begin{align}
\gamma(\epsilon)&=\frac{1}{V}\sum_\vec{k}\frac{f^0(\vec{k})}{\epsilon(\vec{k})-\epsilon}=\frac{1}{8\pi^3}\int
d^3k'\frac{f^0(\vec{k}')}{\epsilon(\vec{k}')-\epsilon}=\frac{1}{2\pi^2}\int dk' k'^2\frac{f^0(k')}{\epsilon(k')-\epsilon}\nonumber\\
&=\frac{m}{\pi^2\hbar^2}\int dk' \frac{\epsilon(k')}{\epsilon(k')-\epsilon}f^0(k')=\frac{m}{\pi^2\hbar^2}\int dk' f^0(k')+\frac{m\epsilon}{\pi^2\hbar^2}\int
dk' \frac{f^0(k')}{\epsilon(k')-\epsilon}.
\end{align}
The first integral can be given by the Sommerfeld expansion
Eq.~(\ref{Eq(C-2):Sommerfeld1}).
\begin{equation}
\frac{m}{\pi^2\hbar^2}\int dk' f^0(k')=\frac{m\sqrt{m}}{\pi^2\hbar^3}\int d\epsilon' \frac{1}{\sqrt{2\epsilon'}} f^0(\epsilon')=2N_0.
\end{equation}
To perform the second integral,
\begin{equation}
\frac{m\epsilon}{\pi^2\hbar^2}\int dk' \frac{f^0(k')}{\epsilon(k')-\epsilon}=\frac{2m^2\epsilon}{\pi^2\hbar^4}\int dk'
\frac{f^0(k')}{k'^2-k^2}=-\frac{m^2\epsilon}{\pi^2\hbar^4 k}\int dk'\ln\left|\frac{k-k'}{k+k'}\right|\partial_{k'}f^0=-\frac{mk}{2\pi^2\hbar^2}\int
d\epsilon'\ln\left|\frac{k-k'}{k+k'}\right|\partial_{\epsilon'}f^0.
\end{equation}
We are now ready to perform the integral in Eq.~(\ref{Eq(C-3):Kondo integral}).

\begin{align}
\int d\epsilon G(\epsilon)\gamma(\epsilon)\partial_\epsilon f^0&=2N_0\int d\epsilon G(\epsilon)\partial_\epsilon f^0-\frac{m}{2\pi^2\hbar^2}\int d\epsilon
d\epsilon' kG(\epsilon)\ln\left|\frac{k-k'}{k+k'}\right|\partial_\epsilon f^0\partial_{\epsilon'}f^0\nonumber\\
&=2N_0\int d\epsilon G(\epsilon)\partial_\epsilon f^0-\frac{m\sqrt{m}}{\sqrt{2}\pi^2\hbar^3}\int d\epsilon d\epsilon'
\sqrt{\epsilon}G(\epsilon)\ln\left|\frac{\sqrt{\epsilon}-\sqrt{\epsilon}'}{\sqrt{\epsilon}+\sqrt{\epsilon}'}\right|\partial_\epsilon\frac{1}{1+e^{\beta(\epsilon-\mu^0)}}\partial_{\epsilon'}\frac{1}{1+e^{\beta(\epsilon'-\mu^0)}}.
\end{align}
The first integral is given by the Sommerfeld expansion
Eq.~(\ref{Eq(C-2):Sommerfeld2}). For the second term, exact
substitution of $\epsilon=\epsilon'$ yields divergence, however, we may still calculate the temperature dependence of the term. By substituting
$X=\beta(\epsilon-\mu^0)$ and $X'=\beta(\epsilon'-\mu^0)$, the second term is
\begin{align}
&-\frac{m\sqrt{m}}{\sqrt{2}\pi^2\hbar^3}\int d\epsilon d\epsilon'
\sqrt{\epsilon}G(\epsilon)\ln\left|\frac{\sqrt{\epsilon}-\sqrt{\epsilon}'}{\sqrt{\epsilon}+\sqrt{\epsilon}'}\right|\partial_\epsilon\frac{1}{1+e^{\beta(\epsilon-\mu^0)}}\partial_{\epsilon'}\frac{1}{1+e^{\beta(\epsilon'-\mu^0)}}\nonumber\\
&=-\frac{m\sqrt{m}}{\sqrt{2}\pi^2\hbar^3}\int dX dX'
\sqrt{k_BTX+\mu^0}G(k_BTX+\mu^0)\ln\left|\frac{\sqrt{k_BTX+\mu^0}-\sqrt{k_BTX'+\mu^0}}{\sqrt{k_BTX+\mu^0}+\sqrt{k_BTX'+\mu^0}}\right|\partial_X\frac{1}{1+e^{X}}\partial_{X'}\frac{1}{1+e^{X'}}\nonumber\\
&\approx-\frac{m\sqrt{m}}{\sqrt{2}\pi^2\hbar^3}\sqrt{\mu^0}G(\mu^0)\int dX dX' \left(\ln \frac{k_BT}{\mu^0}+\ln
\frac{X-X'}{4}\right)\partial_{X}\frac{1}{1+e^{X}}\partial_{X'}\frac{1}{1+e^{X'}}.
\end{align}
Here we expanded with respect to $k_BT$, which is small compared to $\mu^0$.
We drop the second contribution $\ln(X-X')/4$ since it gives a much smaller
contribution than the $\ln k_BT/\mu^0$ contribution at low temperature. Thus
we keep only the logarithmic term.
\begin{equation}
-\frac{m\sqrt{m}}{\sqrt{2}\pi^2\hbar^3}\int d\epsilon d\epsilon'
\sqrt{\epsilon}G(\epsilon)\ln\left|\frac{\sqrt{\epsilon}-\sqrt{\epsilon}'}{\sqrt{\epsilon}+\sqrt{\epsilon}'}\right|\partial_{\epsilon}\frac{1}{1+e^{\beta(\epsilon-\mu^0)}}\partial_{\epsilon'}\frac{1}{1+e^{\beta(\epsilon'-\mu^0)}}\approx-\frac{m}{2\pi^2\hbar^2}k^FG(\mu^0)\ln\frac{k_BT}{\mu^0}.
\end{equation}
By using $N_0=mk^F/2\pi^2\hbar^2$, for low temperature, the following
replacement is valid under an energy integration.
\begin{equation}
\gamma(\epsilon)\partial_\epsilon f^0\to -N_0\ln\frac{k_BT}{\mu^0}\delta(\epsilon-\mu^0),
\end{equation}
which is Eq.~(\ref{Eq:Kondo integral}).
\end{widetext}

\section{Error analysis\label{Appendix:error}}

Errors in the parameters $l_N^{\rm sf}$, $\alpha_{\rm eff}$, $\rho_N$, $\rho_F$ (experimentally determined) and $\kappa$ (determined through a normalization procedure), as well as uncertainty from our fitting method are our main concern in establishing uncertainty in the extracted values of $\eta$. All other parameters are constrained through the previous work, and potential errors in such quantities are not considered.

Although both $\rho_N$ and $\rho_F$ are measured directly from $R_N$ and $R_F$, with very small random noise, these quantities suffer from experimental uncertainty in the wire cross sectional area (through $\rho = RA/L$), particularly a non-rectangular shape. This uncertainty is random between devices, but \textit{systematic} across all $T$ within a single device. It is estimated at the level of $\approx$5 \% from SEM images of the wire edge profile. In our measurement of $\Delta R_{\rm NL}$ we observe a baseline noise floor of around 1 nV (at a modulation frequency of 13 Hz). This corresponds to $\approx3$ $\mu \Omega$ in our measurements and is an absolute noise source independent of signal size.
When fitting  $\Delta R_{\rm NL}(L)$ at each temperature to extract $l_N^{\rm sf}$ and $\alpha_{\rm eff}$, the uncertainties in $\rho$ and $\Delta R_{\rm NL}$ are used to weight a least means square minimization fit, with estimated errors in $l_N^{\rm sf}$ and $\alpha_{\rm eff}$ arising from combining these errors with the fitting residuals. In reality, the parameters $l_N^{\rm sf}$ and $\alpha_{\rm eff}$, are limited in precision by the relative uncertainty in the cross-sectional area measurements, and are therefore largely independent of $T$. Although we may obtain $\alpha_{\rm eff}$ and $l_N^{\rm sf}$ by fitting $\Delta R_{\rm NL}(L,T)$, using a literature value of $l_N^{\rm sf}$ as a constraint on the signal magnitude, the magnitude of $\alpha_{\rm eff}$ is poorly constrained, due to the inherent difficulty in precisely measuring the ferromagnet/normal metal interface resistance. Consequently, errors that are independent of $T$ dominate the extracted values of $\alpha_{\rm eff}$.

With the errors for $l_N^{\rm sf}$ and $\alpha_{\rm eff}$ established, it remains to estimate the uncertainty in $\kappa$, before determining $\eta$. As both $\alpha_{\rm FM}$ and $\alpha_{\rm eff}$ are broadly of a similar magnitude, and $\kappa=\alpha_{\rm eff}/\alpha_{\rm FM}$, the systematic errors in each quantity could, in principle, give an error larger than the estimated value of $1-\kappa$ (typically $1-\kappa$ is around 10 \%, while errors in $\alpha$ are around 5 \% to 10 \%). However, this systematic error is irrelevant for the normalization procedure we use, and is one of the key advantages to our method: As any error in $\alpha$ are largely $T$-independent (errors from both fitting and estimates of interface resistance), they make no impact on the overall normalization factor [$r$ in Eq.~(\ref{Eq(B):normalization})], since any systematic error in $\alpha_{\rm eff}$ or $\alpha_{\rm FM}$ is intrinsically compensated by $r$. Thus we can estimate the error in $\kappa$ solely from the uncertainty in the normalization procedure.  To obtain this value, we realize that the process of minimizing the standard deviation of $\rho_K^1d^1 /\rho_K^id^i$  (our normalization procedure) is identical to a linear regression of $y^i=A^i(1-r^i\kappa^i)$, where $y^i=  \rho_K^1 /\rho_N^i  [1/l_N^{\rm sf,i}-1/(l_N^{\rm sf,i}+\rho_N^i l_F^{\rm sf}[(1-{\alpha_{\rm FM}}^2 ) \rho_F ]^{-1})]$, and $A^i=C_{\rm Fe}^1 d^1\tau_N^{sf}/2C_{\rm Fe}^i d^i\eta \tau_N$. Once again, we use the superscript $i$ to denotes a given dataset (i.e. a given $T_A$), with $i=1$ representing the unannealed data (which can be scaled exactly, see Appendix~\ref{Sec-A:Scaling}). As we can establish both $y$ and $\kappa$ for a given $T_A$, we may therefore estimate the uncertainty in $r$, and so the relative error in our scaled $\kappa$, from the residuals of a least mean squares fit of $y=A(1-r\kappa)$ with $A$ and $r$ as free variables. These estimates are the error bars shown in Fig~\ref{Fig:kappa}.

The final challenge is to incorporate all these errors together for our final fitting procedure to estimate $\eta$. $\eta$ is found from fitting $\eta\rho_K$ from experimental data using our models for $\rho_K$, i.e. through rearranging Eq.~(\ref{Eq:Kappa3}). Through the discussed procedure we now have estimates for all parameters in this equation, including errors for the experimentally determined quantities ($\kappa$, $\rho_N$, $\rho_F$, $\alpha$, $l_N^{\rm sf}$). To establish the error on the experimental $\eta\rho_K$ we use Monte Carlo sampling assuming Gaussian distributed uncertainties for all quantities (\textit{via} the NIST uncertainty machine~\cite{NIST-UM}) to account for the combination of all uncertainties in Eq.~(\ref{Eq:Kappa3}). The extracted errors are subsequently used as weightings for fitting $\eta\rho_K$ to either the G-G or Kondo model, again using a least-mean-square approach. The extracted parameter uncertainties are shown in Fig.~\ref{Fig:kappa} for each method, with the final errors for $\eta$ in panel (c). All quoted errors are a single standard deviation, including those shown for the extracted values of $T_K$.
Most errors are relative rather than absolute, and so data at large $T_A$ appear more error prone than those at low $T_A$, despite the larger $\kappa$. In calculating $\hat{\eta}$ an unweighted average is taken, with the uncertainty in this case quoted as the standard deviation in the eight values.

\end{appendix}


\begin{thebibliography}{99}
\bibitem{Bass07JPCM}%
    J. Bass and W. P. Pratt, %
    J. Phys. Condens. Matter \textbf{19}, 183201 (2007).
\bibitem{Uchida08Nat}%
	K. Uchida, S. Takanashi, K. Harii, J. Ieda, W. Koshibae, K. Ando, S. Maekawa, and E. Saitoh %
	Nature \textbf{455}, 778 (2008).
\bibitem{Valenzuela06Nat}%
	S. O. Valenzuela and M. Tinkham, %
	Nature \textbf{442}, 176 (2006).
\bibitem{Tserkovnyak05RMP}%
	Y. Tserkovnyak, A. Brataas, G. E. W. Bauer, and B. Halperin, %
	Rev. Mod. Phys \textbf{77}, 1375 (2005).
\bibitem{Johnson85PRL}%
    M. Johnson and R. H. Silsbee, %
    Phys. Rev. Lett. \textbf{55}, 1790 (1985).
\bibitem{Jedema01Nat}%
    F. J. Jedema, A. T. Filip, and B. J. van Wees, %
    Nature \textbf{410}, 345 (2001).
\bibitem{Garzon05PRL}%
    S. Garzon, I. \v{Z}uti\'{c}, and R. A. Webb, %
    Phys. Rev. Lett. \textbf{94}, 176601 (2005).
\bibitem{Godfrey06PRL}%
    R. Godfrey and M. Johnson, %
    Phys. Rev. Lett. \textbf{96}, 136601 (2006).
\bibitem{Kimura07PRL}%
    T. Kimura and Y. Otani, %
    Phys. Rev. Lett. \textbf{99}, 196604 (2007).
\bibitem{Yang08NP}%
    T. Yang, T. Kimura, and Y. Otani, %
    Nat. Phys. \textbf{4}, 851 (2008).
\bibitem{Bakker10PRL}%
    F. L. Bakker, A. Slachter, J.-P. Adam, and B. J. van Wees, %
    Phys. Rev. Lett. \textbf{105}, 136601 (2010).
\bibitem{Mihajlovic10PRL}%
    G. Mihajlovi\'{c}, J. E. Pearson, S. D. Bader, and A. Hoffmann, %
    Phys. Rev. Lett. \textbf{104}, 237202 (2010).
\bibitem{Slachter10NP}%
    A. Slachter, F. L. Bakker, J.-P. Adam, and B. van Wees, %
    Nat. Phys. \textbf{6}, 879 (2010).
\bibitem{Fukuma11NM}%
    Y. Fukuma, L. Wang, H. Idzuchi, S. Takahashi, S. Maekawa, and Y. Otani,
    Nat. Mater. \textbf{10}, 527 (2011).
\bibitem{Niimi13PRL}%
    Y. Niimi, D. Wei, H. Idzuchi, T. Wakamura, T. Kato, and Y. C. Otani, %
    Phys. Rev. Lett. \textbf{110}, 016805 (2013).
\bibitem{Gijs97AIP}%
	M. A. M. Gijs and G. E. W. Bauer, %
	Adv. Phys. \textbf{46}, 285 (1997).
\bibitem{Valet93PRB}%
    T. Valet and A. Fert, %
    Phys. Rev. B \textbf{48}, 7099 (1993).
\bibitem{Elliott54PR}%
    R. J. Elliott, %
    Phys. Rev. \textbf{96}, 266 (1954).
\bibitem{Andreev59JETP}%
    V. V. Andreev and V. I. Gerasimenko, %
    Sov. Phys. JETP \textbf{35}, 846 (1959).
\bibitem{Yafet63Book}%
    Y. Yafet, in \textit{Solid State Physics}, %
    edited by F. Seitz and D. Turnbull (Academic, New York, 1963), Vol. 14.
\bibitem{Kimura08PRL}%
    T. Kimura, T. Sato, and Y. Otani, %
    Phys. Rev. Lett. \textbf{100}, 066602 (2008).
\bibitem{Casanova09PRB}%
    F. Casanova, A. Sharoni, M. Erekhinsky, and I. K. Schuller, %
    Phys. Rev. B \textbf{79}, 184415 (2009).
\bibitem{Otani11PTAMPES}%
    Y. Otani and T. Kimura, %
    Philos. Trans. A. Math. Phys. Eng. Sci. \textbf{369}, 3136 (2011).
\bibitem{Erekhinsky12APL}%
    M. Erekhinsky, F. Casanova, I. K. Schuller, and A. Sharoni, %
    Appl. Phys. Lett. \textbf{100}, 212401 (2012).
\bibitem{Kimura12PRB}%
    T. Kimura, N. Hashimoto, S. Yamada, M. Miyao, and K. Hamaya, %
    NPG Asia Mater. \textbf{4}, e9 (2012).
\bibitem{Zou12APL}%
    H. Zou and Y. Ji, %
    Appl. Phys. Lett. \textbf{101}, 082401 (2012).
\bibitem{Villamor13PRB}%
    E. Villamor, M. Isasa, L. E. Hueso, and F. Casanova, %
    Phys. Rev. B \textbf{87}, 094417 (2013).
\bibitem{Batley15PRB}%
    J. T. Batley, M. C. Rosamond, M. Ali, E. H. Linfield, G. Burnell, and B. J. Hickey, %
    Phys. Rev. B \textbf{92}, 220420(R) (2015).
\bibitem{Brien14NC}%
    L. O'Brien, M. J. Erickson, D. Spivak, H.Ambaye, R. J. Goyette, V.Lauter, P. A. Crowell, and C. Leighton, %
    Nat.Commun. \textbf{5}, 3927 (2014).
\bibitem{Brien16PRB}%
    L. O'Brien, D. Spivak, J. S. Jeong, K. A. Mkhoyan, P. A. Crowell, and C. Leighton, %
    Phys. Rev. B \textbf{93}, 014413 (2016).
\bibitem{Kondo64PTP}%
    J. Kondo, %
    Prog. Theor. Phys. \textbf{32}, 37 (1964).
\bibitem{Domenicali61JAP}%
	C. Domenicali and E. Christenson, %
	J. Appl. Phys. \textbf{32}, 2450 (1961).
\bibitem{Taniyama16}%
	K. Hamaya, T. Kurokawa, S. Oki, S. Yamada, T. Kanashima, and T. Taniyama,
    arXiv:1603.08599.
\bibitem{Taniyama03PRL}%
	T. Taniyama, N. Fujiwara, Y. Kitamoto, and Y. Yamazaki,
	Phys. Rev. Lett., \textbf{90}, 016601 (2003)
\bibitem{Fert95PRB}%
    A. Fert, J.-L. Duvail, and T. Valet, %
    Phys. Rev. B \textbf{52}, 6513 (1995).
\bibitem{Goldhaber-Gordon98PRL}%
    D. Goldhaber-Gordon, J. G\"{o}res, and M. A. Kastner, H. Shtrikman, D. Mahalu, and U. Meirav,%
    Phys. Rev. Lett. \textbf{81}, 5225 (1998).
\bibitem{vanderWiel00Sci}%
    W. G. van der Wiel, S. De Franceschi, T. Fujisawa, J. M. Elzerman, S. Tarucha, and L. P. Kouwenhoven, %
    Science \textbf{289}, 2105 (2000).
\bibitem{Parks07PRL}%
    J. J. Parks, A. R. Champagne, G. R. Hutchison, S. Flores-Torres, H. D. Abruna, and D. C. Ralph, %
    Phys. Rev. Lett. \textbf{99}, 026601 (2007).
\bibitem{Takahashi03PRB}%
	S. Takahashi and S. Maekawa, %
	Phys. Rev. B \textbf{67} 052409 (2003).
\bibitem{Huchinson51PR}%
	T. Hutchison and J. Reekie, %
	Phys. Rev. \textbf{83}, 854 (1951).
\bibitem{Mydosh93}%
	J. Mydosh, in \textit{Spin Glasses: an Experimental Introduction}, %
	(Taylor and Francis, London, 1993).
\bibitem{Beuneu78PRB} %
	F. Beuneu and P. Monod, %
	Phys. Rev. B \textbf{18}, 2422 (1978).
\bibitem{Ashcroft76}%
	N. W. Ashcroft and N. D. Mermin, in \textit{Solid State Physics}, %
	(Holt, Rinehart, and Winston, 1976).
\bibitem{Monod67PRL}%
	P. Monod, %
	Phys. Rev. Lett. \textbf{19}, 1113 (1967).
\bibitem{Idzuchi12APL}%
	H. Idzuchi, Y. Fukuma, L. Wang and Y. Otani, %
	App. Phys. Lett. \textbf{101}, 022415 (2012).
\bibitem{NIST-UM}%
	T. Lafarge and A. Possolo, %
    NCLSI Measure Journal of Measurement Science \textbf{10}, 20 (2015); %
    see \url{http://uncertainty.nist.gov}
\end{thebibliography}
\end{document}